\documentclass{ws-ijmpb}

\usepackage{graphicx}
\usepackage{dcolumn}
\usepackage{bm}
\usepackage{bbm}
\usepackage{psfrag}
\usepackage{stmaryrd}
\usepackage{subfigure}
\usepackage[sort&compress,super,comma]{natbib}

\newcommand{\refsec}[1]{Sect.~\ref{#1}}
\newcommand{\reffig}[1]{Fig.~\ref{#1}}
\newcommand*{\text}[1]{\mbox{#1}}
\newcommand*{\ind}[1]{\mbox{\scriptsize #1}}
\newcommand{\I}{\textup{i}}
\newcommand{\Iexp}{\mbox{\scriptsize i}}
\newcommand{\E}{\textup{e}}
\newcommand{\D}{\textup{d}}
\newcommand{\dd}{\mbox{d}}
\newcommand{\dod}[2]{\frac{\dd #1}{\dd #2}}

\newcommand{\pop}[2]{\frac{\partial #1}{\partial #2}}
\newcommand{\poptxt}[2]{\partial #1 / \partial #2}

\newcommand*{\laplace}{\Delta}
\newcommand{\tr}[2][]{\mbox{Tr}_{#1}\left\{#2\right\}}
\newcommand{\trtxt}[2][]{\mbox{Tr}_{#1}\{#2\}}
\newcommand*{\vek}[1]{\bm{#1}}
\newcommand*{\tens}[1]{\mathsf{#1}}
\newcommand{\ganz}{\mathbbm{Z}}
\newcommand*{\ketL}[1]{\mathopen{|}#1\mathclose{)}}
\newcommand*{\braL}[1]{\mathopen{(}#1\mathclose{|}}
\newcommand*{\sprodL}[2]{\mathopen{(}#1 |#2 \mathclose{)}}
\newcommand*{\ketbraL}[2]{\mathopen{|}#1\mathclose{)}\mathopen{(}#2\mathclose{|}}
\newcommand*{\opL}[1]{\hat{\mathfrak{#1}}}

\newcommand*{\bra}[1]{\mathopen{\langle}#1\mathclose{|}}
\newcommand*{\ket}[1]{\mathopen{|}#1\mathclose{\rangle}}

\newcommand*{\hamil}{\hat{H}}

\newcommand*{\dens}{\hat{\rho}}
\newcommand*{\Paulix}{\hat{\sigma}_x}
\newcommand*{\Pauliy}{\hat{\sigma}_y}
\newcommand*{\Pauliz}{\hat{\sigma}_z}

\newcommand*{\Paulipl}{\hat{\sigma}_+}
\newcommand*{\Paulimi}{\hat{\sigma}_-}

\newcommand*{\Kommutatortxt}[3][]{[#2,#3]_{#1}}
\newcommand*{\Haver}[1]{\mathopen{\llbracket} #1 \mathclose{\rrbracket}}

\newcommand*{\Hspace}{\mathcal{H}}
\newcommand*{\Lspace}{\mathcal{L}}

\begin{document}

\markboth{M.~Michel, J.~Gemmer \& G.~Mahler}
{Microscopic Quantum Mechanical Foundation of Fourier's Law}

%
\catchline{}{}{}{}{}
%

\title{MICROSCOPIC QUANTUM MECHANICAL FOUNDATION OF FOURIER'S LAW}

\author{MATHIAS MICHEL}

\address{Institut f\"ur Theoretische Physik I, Universit\"at Stuttgart, 
Pfaffenwaldring 57\\
D-70550 Stuttgart, Germany\\
mathias.michel@itp1.uni-stuttgart.de}

\author{JOCHEN GEMMER}

\address{Fachbereich Phyik, Universit\"at Osnabr\"uck
Barbarastr. 7\\
D-49069 Osnabr\"uck, Germany}

\author{G\"UNTER MAHLER}

\address{Institut f\"ur Theoretische Physik I, Universit\"at Stuttgart, 
Pfaffenwaldring 57\\
D-70550 Stuttgart, Germany}

\maketitle

\begin{history}
\received{31 October 2006}
\end{history}

\begin{abstract}
Besides the growing interest in old concepts like temperature and entropy at the nanoscale, theories of relaxation and transport have recently regained a lot of attention.
With the electronic circuits and computer chips getting smaller and smaller, a fresh look should be appropriate on the equilibrium and nonequilibrium thermodynamics at small length scales far below the thermodynamic limit,
 i.e.\ on the theoretical understanding of original macroscopic processes, e.g.\ transport of energy, heat,
 charge, mass, magnetization etc.
Only from the foundations of a theory its limits of applicability may be inferred.
This review tries to give an overview of the background and recent developments in the field of nonequilibrium quantum thermodynamics focusing on the transport of heat in small quantum systems. 
\end{abstract}


%
%
\section{Introduction}

How is energy or heat transported through a material system on a microscopic level of description?
Despite its long history this old question has remained exciting even today, since a satisfactory answer has not yet been found.
The large number of different publications on the topic of \emph{heat transport} indicates both an increasing interest in the problem and its nontrivial character, emphasized by the title of a recent publication by Bonetto\cite{Bonetto2000} et al.\ ``Fourier's Law: A Challenge to Theorists''.

\newpage

The review at hand tries to give an introduction to the phenomenon of transport of heat from the viewpoint of various theoretical approaches, before turning to some new perspectives on this old topic.
In such nonequilibrium phenomena as heat conduction, the intensive parameters of different parts of a system are no longer equal, i.e., there is no global equilibrium established inside the whole system.
However, despite of this fact, in a \emph{macroscopically small} but \emph{microscopically large}\cite{Lepri2003} part of the system, at least near the global equilibrium, one finds an equilibrium state slightly different from region to region.  
One may think, for example, of a material bar coupled at both ends to heat baths of different temperature.
As noted by Fourier\cite{Fourier1955}
\begin{quote}
When heat is unequally distributed among the different parts of a solid mass, it tends to attain equilibrium, and passes slowly from the parts which are more heated to those which are less; and at the same time it is dissipated at the surface, and lost in the medium or in the void.
\end{quote}
In this sort of experiment one typically finds a constant temperature gradient within the material.
Thus, there is no global temperature defined in the whole system, but in a small enough part one could approximately describe the system by a canonical equilibrium state with \emph{local} temperature.

The first detailed theoretical as well as experimental investigation of heat transport through solids was carried out by Joseph Fourier\cite{Fourier1955} about 1807.
The subsequent Paragraph is dedicated to him introducing the famous \emph{Fourier's Law}.
In the following we will review some of the main approaches to the famous law on microscopic level as well as their respective shortcomings, before we give an overview of recent developments in the heat conduction research.
The second Section will introduce some main ingredients of nonequilibrium physics followed by the analysis of heat conduction in small quantum systems by the quantum master equation.
The last part of this review presents a recently developed theory, the so-called Hilbert Space Average Method (HAM), to study the relaxation into equilibrium.
This method allows to approach thermal transport from a completely different viewpoint.

%
%
\section{Overview}

%
%
\subsection{Fourier's Law}
\label{chap:HeatTransport:sec:FLaw}

Almost two hundred years ago Fourier conjectured that temperature (or as we know today: energy) tends to diffuse\cite{Fourier1955} through solids once close enough to equilibrium.
In his work he considered the transport of heat on a phenomenological respectively macroscopic level, defining a partial differential equation for the heat transport through some material and trying to solve it by mathematical methods, e.g., the Fourier series expansion.
Nowadays, Fourier is mainly known for his marvelous mathematical research, especially the celebrated Fourier analysis.

Considering a situation with a position and time dependent temperature profile $T(\vek{r},t)$ within some material, Fourier describes the problem by an equation for the internal energy density $u(T(\vek{r},t))$.
The change of this quantity $u(T(\vek{r},t))$ should be proportional to the gradient of the thermodynamic force -- here the temperature gradient $\vek{\nabla} T(\vek{r},t)$,
\begin{equation}
  \label{eq:1}
  \pop{}{t} u\big(T(\vek{r},t)\big)
  = \kappa\vek{\nabla}\cdot\vek{\nabla} T(\vek{r},t)\;,
\end{equation}
with $\kappa$ being the heat conductivity.
As the energy density is a function of temperature one can also state that
\begin{equation}
  \label{eq:2}
  \pop{}{t} u\big(T(\vek{r},t)\big)
  = \pop{u}{T}\pop{T}{t} 
  = \kappa\laplace T(\vek{r},t)\;,
\end{equation}
and by using the definition of the specific heat capacity 
\begin{equation}
  \label{eq:3}
  c(T)=\pop{u}{T}
\end{equation}
one finds the \emph{heat conduction equation}
\begin{equation}
  \label{eq:1025}
  \pop{}{t}T(\vek{r},t) = \frac{\kappa}{c(T)} \laplace T(\vek{r},t)\;.
\end{equation}
This diffusive behavior is slightly reformulated by applying the continuity equation for the energy density 
\begin{equation}
  \label{eq:26}
  \pop{u}{t}+\vek{\nabla}\cdot\vek{J}=0\;,
\end{equation}
where $\vek{J}$ represents the heat current in the system.
 Plugging (\ref{eq:26}) into (\ref{eq:2}) one finds
\begin{equation}
  \label{eq:27}
  \vek{\nabla}\cdot\vek{J} 
  = - \vek{\nabla}\cdot\Big(\kappa\vek{\nabla}T(\vek{r},t)\Big)
  \;.
\end{equation}
From these considerations one immediately gets what \emph{Fourier's law} states close to thermal equilibrium: a proportionality between the heat current and the temperature gradient
\begin{equation}
  \label{eq:28}
  \vek{J} = - \kappa\vek{\nabla}T(\vek{r},t)\;.
\end{equation}
Furthermore by thinking of the temperature $T$ being a function of the internal energy density $u$ we could state for the energy current
\begin{equation}
  \label{eq:29}
  \vek{J} 
  = - \kappa \vek{\nabla} T\big(u(\vek{r},t)\big)
  = - \kappa \pop{T}{u} \vek{\nabla} u
  = - \frac{\kappa}{c}\vek{\nabla} u(\vek{r},t)\;,  
\end{equation}
where $\vek{J}$ is now the energy current density.

Despite the ubiquitous occurrence of the phenomenon of heat conduction in our everyday experience, its explanation on the basis of some reversible microscopic dynamics remains a serious problem.
The quantity of interest here is the material property $\kappa$ -- the heat conductivity.
In the last two centuries there have been several different microscopic approaches to Fourier's Law from completely different backgrounds.
The goal of all of those efforts was to give a thorough and comprehensive picture of heat transport from the viewpoint of a basic microscopic theory.

%
%
\subsection{Kinetic Gas Theory for Phonons}
\label{chap:HeatTransport:sec:debye}

One of the first attempts to find a microscopic foundation for the heat conductivity of an insulating solid was due to Debye\cite{Debye1914,Kittel1969}.
He considered the phonons inside such a material as classical particles using results of the kinetic gas theory to describe the heat transport based on phonons.

The particle current in one direction of a crystal should be proportional to the density $\eta$ of the particles and their mean velocity $\overline{v}$.
A single phonon considered as a classical particle moving from a region with temperature $T+\Delta T$ to a region with temperature $T$ looses the energy $C\Delta T$ with $C$ being the heat capacity of the respective particle.
With the internal temperature gradient $\vek{\nabla}T$ along the free path $\overline{l}$ of the phonon we find the temperature difference to be
\begin{equation}
  \label{eq:30}
  \Delta T = \overline{l}\,\vek{\nabla}T\;.
\end{equation}
In such a situation the energy current is just proportional to the particle current times the transported energy and thus
\begin{equation}
  \label{eq:31}
  \vek{J} \propto \eta\,\overline{v}\, C\Delta T 
          \propto \eta\,\overline{v}\, C\,\overline{l}\,\vek{\nabla}T
  \;.
\end{equation}
Reformulating this result by the heat capacity per volume $c=\eta C$ and comparing it to Fourier's Law (\ref{eq:28}) we find for the heat conductivity coefficient
\begin{equation}
  \label{eq:32}
  \kappa \propto c\,\overline{v}\,\overline{l}\;.
\end{equation}

As the free path of the phonons enters the above microscopic formula an infinite heat conductivity would result in case of no scattering in the solid.
Why should phonons have a finite free path inside a material?
Debye\cite{Debye1914} and others have mainly taken two processes into account which limit the free path of the phonons, the scattering at impurities (primarily in amorphous materials) and the phonon-phonon interaction in the solid (perfect crystal, lacking any impurities).
However, if the lattice was harmonic there would be no scattering between phonons and therefore no finite heat conductivity.

To include anharmonicity in the coupling of the atoms in the solid is a serious theoretical problem.
Debye\cite{Debye1914} proposed that the Brownian motion of the atoms might produce an irregularity in the lattice and thus give rise to phonon scattering in an otherwise perfect crystal.
On this basis he was able to account for the mean free path of the phonons, finding it anti-proportional to the temperature of the solid.
In this way he was able to derive an approximate formula for a finite heat conductivity\cite{Debye1914} from a microscopic point of view.

%
%
\subsection{Peierls-Boltzmann Equation}
\label{chap:HeatTransport:sec:peierls}

A more detailed approach to the question of a microscopic foundation of Fourier's Law was introduced by Peierls\cite{Peierls1929,Peierls1955} in 1929 applying the Boltzmann equation to the phonons in the solid.
Peierls mainly considered the mentioned anharmonicities in the atom coupling, which are the dominant effect underlying phonon scattering in pure crystals of not too small size and at not too low temperatures.

Treating the anharmonic part of the interaction as a perturbation, Peierls was able to find conditions from Fermi's Golden Rule for the possible scattering processes: the energy conservation and the momentum conservation
\begin{eqnarray}
  \label{eq:33}
  \omega_1+\omega_2 &=& \omega_3\;,\\
  \label{eq:1034}
  \vek{k}_1+\vek{k}_2 &=& \vek{k}_3+ n \vek{g}\;,\quad n\in\ganz\;,
\end{eqnarray}
where $(\omega_1, \vek{k}_1)$ and $(\omega_2,\vek{k}_2)$ are the two incoming phonons, $(\omega_3,\vek{k}_3)$ is the outgoing one, and $\vek{g}$ is the appropriate reciprocal lattice vector\cite{Kittel1969}.
All processes with $\vek{g}=0$ are not suitable for a thermalization of the phonon gas, since then the whole momentum of all phonons is conserved, the conductivity would again be infinite (all these processes are called \emph{normal} or N-processes).
Of vital importance for the existence of a thermal resistance are the scattering processes with $\vek{g}\neq0$ called \emph{Umklapprozess}.

Considering now very low temperatures ($T<\Theta$, $\Theta$ Debye temperature) where most of the two incoming thermally activated phonons have not enough energy to fulfill the above scattering conditions, there is again no possible umklapprocess.
According to the Boltzmann distribution of phonon energies we expect that for low temperatures the number of high-energy phonons decreases by an exponential $\E^{-\Theta/T}$.
Thus, the conductivity must be proportional to that exponential factor in the low temperature regime.

Since phonons are also scattered at impurities or boundaries of the crystal, one has to consider more scattering processes in order to get a complete picture of heat conduction in solids\cite{Michel1976,Michel1978,Michel1978II,Ziman2001,Kittel1969}.
These additional processes are very important for low temperatures, especially, since in this region the umklapprocesses die out rapidly.

To describe the phonons inside the crystal Peierls essentially proposed a Boltzmann equation replacing classical gas particles by quantized quasi particles -- the phonons.
Furthermore he assumed classical transition probabilities between different phonon modes as obtained from Fermi's Golden Rule.
His theory succeeds in explaining the basic transport properties of solids and allows to derive statements about the conductivity of insulating crystals.
The equation itself is very hard to solve, but, nevertheless, one can obtain some tendencies in the temperature dependence of the conductivity.
Peierls\cite{Peierls1955} states:
\begin{quote}
The only prediction that would seem to follow with certainty is that the law must be intermediate between $T^{-1}$ and $T^{-2}$.
\end{quote}
Here the different dependencies between $1/T$ and $1/T^2$ follow from third order respectively forth order anharmonicities in the interaction.

However, the Peierls-Boltzmann theory faces conceptual shortcomings: quantized normal modes of a many particle system are described classically, i.e., they are well localized in configuration as well as momentum space and show no dispersion\cite{Peierls1955}. 
This becomes a serious problem at least in the limit of systems consisting of only a few particles at all.
Already Debye's approach to heat conduction deals with a rather classical idea of the quasi-particles -- thinking of bouncing balls like gas particles in a box instead of quantum mechanical objects.
Furthermore, in order to exploit Fermi's Golden Rule, the actual quantum state of a phonon mode is discarded keeping only the mean occupation number. 
Due to the neglect of any phases, this is called the random phase approximation and replaces Boltzmann's Stosszahlansatz (see Peierls\cite{Peierls1955}).
Strictly speaking these both lack justification, though, and there are no clear criteria for the applicability of the theory.
Additionally, the application of the theoretical concept to a concrete coupling model is very complicated and Peierls\cite{Peierls1955} himself states:
\begin{quote}
This seems a poor return for a long discussion, but progress beyond this stage is difficult, unless we could construct a dispersion law which was simple enough to allow us to list the solutions for the possible phonon collisions explicitly, and yet realistic enough to give the right kind of collision including those of the Umklapp type.
\end{quote}
Even today there are doubts about the importance of umklapprocesses\cite{Wagner1999,Fermi1955} at all for a finite conductivity in solid states.

%
%
\subsection{Linear Response Theory}
\label{chap:HeatTransport:sec:kubo}

Another powerful technique within the field of heat transport is the Green-Kubo formula. 
Derived on the basis of linear response theory it has originally been formulated for electrical transport\cite{Kubo1957,Kubo1991,Mori1956}.
Here the external force, the electric field, is considered as the perturbation of the system.
In this context the current is viewed as the response to this external perturbative electrical potential which can be expressed as a part of the Hamiltonian of the system\cite{Kubo1957}.

%
%
\subsubsection{Currents, Forces and Onsager Relation}
\label{chap:HeatTransport:sec:kubo:1}

In a linear response theory, currents $\vek{J}_i$ are linear function of external forces $\vek{X}_j$ defining the matrix of \emph{response coefficients} $\tens{Z}_{ij}$ as
\begin{equation}
  \label{eq:14}
  \vek{J}_i = \sum_j \tens{Z}_{ij} \,\vek{X}_j\;.
\end{equation}
From the Onsager relation we know that the coefficient matrix has to be symmetrical $\tens{Z}_{ij}=\tens{Z}_{ji}$\cite{Kubo1991,Mahan1981,Groot1962}.

The choice for the current operators and the corresponding forces is ambiguous.
But it turns out that the Onsager relation is not valid for all choices. 
A special one for which the Onsager relation is valid was proposed by de Groot\cite{Groot1962}, who chose currents and forces implicitly defined by the expression for the increase of entropy
\begin{equation}
  \label{eq:15}
  \pop{S}{t}=\sum_i \vek{J}_i \vek{X}_i > 0\;,
\end{equation}
since a nonequilibrium process should be associated with a net entropy production.

Let us consider a situation with two different fluxes, a particle current $\vek{J}_1$ due to an external force $\vek{X}_1=-\frac{1}{T}\vek{\nabla}\mu$ with the generalized potential $\mu$ (contains chemical as well as electrical potentials) and a heat current $\vek{J}_2$ due to the force $\vek{X}_2=\vek{\nabla}\frac{1}{T}$.
Then (\ref{eq:14}) reads
\begin{eqnarray}
  \label{eq:16}
  \vek{J}_1 = -\frac{1}{T} \tens{Z}_{11} \vek{\nabla}\mu 
              + \tens{Z}_{12} \vek{\nabla}\frac{1}{T}\;,\\
  \label{eq:17}
  \vek{J}_2 = -\frac{1}{T} \tens{Z}_{21} \vek{\nabla}\mu
              + \tens{Z}_{22} \vek{\nabla}\frac{1}{T}\;.
\end{eqnarray}
For a vanishing particle current $\vek{J}_1=0$ it follows from (\ref{eq:16})
\begin{equation}
  \label{eq:18}
  \frac{1}{T} \tens{Z}_{11} \vek{\nabla}\mu 
  =\tens{Z}_{12} \vek{\nabla}\frac{1}{T}
\end{equation}
and, using Onsager's relation and reformulating (\ref{eq:17}) with (\ref{eq:18}), for the heat current
\begin{equation}
  \label{eq:19}
  \vek{J}_2 = \Big(\tens{Z}_{22}-\frac{(\tens{Z}_{12})^2}{\tens{Z}_{11}}\Big)
              \vek{\nabla}\frac{1}{T}\;.
\end{equation}
Putting aside any further perturbations like chemical or electrical potentials  ($\mu=0$) we obtain 
\begin{equation}
  \label{eq:20}
  \vek{J}_2 = \tens{Z}_{22}\vek{\nabla}\frac{1}{T}
            = \tens{Z}_{22}\left(-\frac{1}{T^2}\vek{\nabla}T\right)\;.
\end{equation}
Comparing this result to Fourier's Law (\ref{eq:28}) one finds for the heat conductivity
\begin{equation}
  \label{eq:21}
  \kappa = \frac{\tens{Z}_{22}}{T^2}\;.
\end{equation}
For a microscopic theory of the heat conductivity it remains to compute the response coefficients $\tens{Z}_{22}$.

%
%
\subsubsection{Transport Coefficient}
\label{chap:HeatTransport:sec:kubo:2}

In case of an electric perturbation of a system Kubo was able to derive a formula for the conductivity tensor $\tens{\sigma}_{ij}(\omega)$ by first order perturbation theory.
In so doing he considered the external electric potential given as an operator part within the Hamiltonian of the system.
The result of this consideration is basically a current-current autocorrelation function\cite{Kubo1957}
\begin{equation}
  \label{eq:22}
  \tens{\sigma}_{ij}(\omega)
  = \int_0^{\infty}\D t\,\E^{\Iexp\omega t} 
    \int_0^{\beta} \D\lambda \,
    \tr{\vek{J}_j(-t-\I\hbar\lambda)\,\vek{J}_i\,\rho_0}
  \;,
\end{equation}
with the frequency $\omega$ of the perturbation and $\rho_0$ being the equilibrium state of the system at inverse temperature $\beta$.
 The current operators are here in the Heisenberg picture, i.e., for a system with an unperturbed Hamiltonian $\hamil$ they are defined as
\begin{equation}
  \label{eq:23}
  \vek{J}(x):=
  \E^{\frac{\Iexp}{\hbar}\hamil x} \vek{J} 
  \E^{-\frac{\Iexp}{\hbar}\hamil x}\;.
\end{equation}

Since the above Kubo formula essentially consists of a current-current auto-correlation, it may ad hoc be transfered to heat transport\cite{Luttinger1964} simply by replacing the electrical current by the heat current. 
However, the justification of this replacement remains a conceptual problem since there is no way of expressing a temperature gradient in terms of an addend to the Hamiltonian which has been no problem for the electric case.
Remarkably enough, Kubo\cite{Kubo1991} himself commented on that replacement in a rather critical way:
\begin{quote}
Therefore, the treatment developed ...[above] does not to be directly applicable to nonequilibrium states produced by such thermal forces, in order to obtain explicit formulas expressing the responses to thermal forces. It is generally accepted, however, that such formulas exist and are of the same form as those for responses to mechanical disturbances.
\end{quote}

Nevertheless, Luttinger\cite{Luttinger1964} proposed in 1964 a formula for the response coefficient for thermal perturbations on the basis of Kubo's theory
\begin{equation}
  \label{eq:24}
  \tens{Z}_{22} 
  = 
    \int_0^{\infty}\D t\,\E^{-s t} 
    \int_0^{\beta}\D\lambda\, 
    \tr{\vek{J}(-t-\I\hbar\lambda)\,\vek{J}\,\rho_0}\;.
\end{equation}
Additionally, one needs the limit of $\omega\rightarrow 0$ here since a frequency dependent thermal perturbation seems even more suspicious than the questionable analogy to electric perturbations.
The exponential function here with $s>0$ is a switch-on function of the perturbation from the infinite past, to guarantee the convergence of the integral.

There are some further discussions about the divergence of the response coefficient for $\omega\rightarrow0$ and the respective weight of the delta-function (Drude-weight) as well as the regular part inside the above formula. 
But these discussions are definitely beyond the scope of this text and therefore we would like to refer to the literature (see e.g.\ the work by Heidrich-Meisner\cite{Heidrich2005}).

Since there is no detailed derivation based on assumptions or preconditions, it is hard to state the limits of applicability of the formula.
Despite all insufficiencies of this approach, it has become a widely employed technique\cite{Zotos1997,Heidrich2003,Kluemper2002} and it allows for a straightforward application to any system, once it is partially diagonalized. 

%
%
\subsection{Survey of Recent Developments}
\label{chap:HeatTransport:sec:recent}

During the past decade the established field of ``heat transport'' has regained attention in the physics community.
This is not only a consequence of the conceptual difficulties of the above described standard microscopic approaches -- the Peierls-Boltzmann theory and the Green-Kubo formula -- but is also a result of some new and, in part, technical developments.

In a time where electronic circuits and computers are getting smaller and smaller, the knowledge about \emph{thermodynamical properties on the nanoscale} becomes more and more important.
Especially, the quantum limits to some standard thermodynamical concepts may eventually lead to a revolution not only in the theoretical and experimental understanding of the dynamics of small systems, but also in their technical applications.
Thus, there is an increasing interest in experimental as well as theoretical advancements concerning heat transport in nanoscopic devices.
All together this has led to a revival of old, but to a certain degree unsolved questions and to a new vital discussion on heat conduction, especially in small quantum systems.

Unfortunately, the microscopic foundation of Fourier's Law seems to be very complicated and far from being trivial as stated by the above mentioned article by Bonetto\cite{Bonetto2000}.
The numerous difficulties tempted the authors to promise a bottle of wine for a proper microscopic theory.
Another statement about the actual status of heat conduction research appeared recently in an article of Buchanan\cite{Buchanan2005} in Nature Physics: ``No one has yet managed to derive Fourier's Law truly from fundamental principles''.
However, we hope to convince the reader of this text that there are now several ideas of how to approach normal heat conduction and thus Fourier's Law from quantum mechanics directly.

The following selection of heat conduction research can hardly do justice to respective approaches nor claims completeness.
It is only meant to show the great variety of different ideas and their links to  our work.

\subsubsection{Heat Conduction in Classical Systems}

Since the work at hand is mainly based on quantum mechanical ideas, we only briefly comment on classical models of heat conduction.
We refer the interested reader to the excellent review article by Lepri\cite{Lepri2003} et~al., which summarizes the central techniques and the main results of heat conduction in classical mechanics.

In the classical domain it seems to be largely accepted that normal transport requires chaotic microscopic dynamics\cite{Casati1984,Prosen1998} whereas ballistic transport is typically to be found in completely integrable systems following a regular dynamics.
Therefore, one finds, e.g., a normal transport in a Lorentz gas\cite{Vollmer2002,Larralde2003} model which is strongly chaotic (of course, the heat transport is strictly connected to particle diffusion, here).
However, there have also been successful attempts to observe normal transport in the absence of exponential instability\cite{Li2004}, i.e.\ without chaos. 

In view of the many different but often strange and unrealistic model systems under consideration one gets the impression that it is really hard to find the desired diffusive behavior. 
This contradicts our all day experience, since normal diffusion clearly is the typical property rather than ballistic behavior.
Nevertheless, people have proposed several ``spring and ball'' models combined with classical free particle models and different masses etc. (``ding-a-ling''-model\cite{Casati1984}, ``ding-a-dong''-model\cite{Prosen1992}) as well as chains of nonlinear oscillators\cite{Hu1998,Garrido2001,Lepri1997,Lepri1998,Savin2002,Dhar2001} to finally obtain normal transport behavior.

\subsubsection{Experimental Investigations}  

Recently, there have been some efforts to measure the heat conductivity of a single carbon nanotube.
Besides some evidences of ballistic phonon transport in carbon nanotubes\cite{Chiu2005,Yu2005}, there are measurements on the length and temperature dependence of the heat conductivity\cite{Fujii2005}, too.

Furthermore, there are several investigations on the thermal conductance of magnetic systems.
Those materials are insulators, for which the heat transport is not dominated by electrons.
Inside such materials one frequently finds chains, ladders or quasi two dimensional spin structures.
Thus, besides the standard phonon induced heat transport in insulators, also magnons could be involved.
A characteristic of such a behavior is the large anisotropy between the thermal conductivity measured parallel to spin structures and perpendicular to them.
Furthermore, the absolute value of the heat conductivity in some insulating compounds is extremely high, and comparable with the conductivity of some metals.
This has led to the assumption of ballistic magnon transport in such materials.
Recent measurements have been carried out by Sologubenko\cite{Sologubenko2000I,Sologubenko2000II,Sologubenko2000III,Sologubenko2003} et~al. and Hess\cite{Hess2001,Hess2004,Ribeiro2005} et~al.
For a nice overview on experimental results in different materials cf.\ the work of Heidrich-Meisner\cite{Heidrich2005}.

\subsubsection{Kubo-Formula}

To support the experimental results mentioned in the last Section there are several attempts to evaluate the thermal Kubo-formula for spin structures.
Inspite of the above discussed insufficiencies of the Kubo-formula, it is, nevertheless, a frequently used technique to account for the thermal conductivity of a system.
Furthermore, it allows for a discussion of the emergence of regular transport, scaling properties of the conductivity and temperature dependencies.
Besides some analytical attempts for integrable systems\cite{Zotos1996,Zotos1997}, there are numerous numerical ones\cite{Heidrich2005,Heidrich2002,Heidrich2003,Heidrich2004II,Heidrich2005II,Kluemper2000,Kluemper2002}, computing the Kubo-formula for several concrete spin systems.
The main quantity is the Drude weight, i.e.\ the conductivity for frequency going to zero.
A divergent weight indicates ballistic transport.
One promising method evaluates the whole frequency dependent conductivity first, a set of delta-peaks with different weight for finite systems.
Sorting those peaks into bins and estimating the value for frequency to zero, results in the desired conductivity\cite{Heidrich2005II,Jung2006}.

The big advantage of the Kubo formula is its computability for a concrete model after having partially diagonalized its Hamiltonian.
However, it comes without any criteria for its applicability, thus, potentially leading to wrong results.
According to a new theory\cite{Gemmer2006II} it seems to be possible to find the celebrated formula for thermal transport by first principles.

\subsubsection{Reservoir Coupling}

Essentially, the Kubo formula is based on a linear response theory.
But, in order to derive the proper equation, the external perturbation has to be formulated as a potential part in the Hamiltonian of the system.
As discussed above this is not possible for thermal perturbations.
To overcome this problem, such Kubo-scenarios have recently been transfered from Hilbert- to Liouville space, where temperature gradients may indeed be formulated in terms of operators\cite{Michel2005}.
In this approach several heat reservoirs are coupled to the system described by a quantum master equation.
Numerically, the method reveals normal heat transport\cite{Saito1996,Saito1996II,Saito2000,Saito2002,Saito2003,Michel2003} in the final stationary local equilibrium state, already in surprisingly small quantum systems. 
Furthermore, there are some attempts to solve such bath scenarios analytically, e.g. for the Ising model\cite{Lecomte2005}.

Another approach based on the coupling the system to different heat baths leads to the celebrated Landauer-B\"uttiker formula\cite{Landauer1957}.
Here, the system is considered as a junction, e.g.\ a small molecule, which transmits or reflects incoming modes from the two reservoirs of different temperature. 
The heat flow is thus defined only by the transmitted modes.
Again the model reveals both normal and ballistic transport in dependence of the type of the junction\cite{Segal2003,Segal2005I,Segal2005II}.
However, since the system does not contain an internal geometrical substructure, there is no room for getting Fourier's Law inside the system proper. 

Instead of the coupling to reservoirs of different temperature in Liouville space, there are proposals to impose a current inside the system by adding the standard current operator\cite{Antal1997,Antal1999,Eisler2003} to the Hamiltonian of the system.
Finally, one observes again the quasi stationary state in the system finding domains of normal respectively ballistic transport in spin chains.

Finally, there are efforts to study the heat conducting behavior of large spin systems by the new numerical method of time dependent density matrix renormalization group\cite{Gobert2005} (tDMRG).

\subsubsection{Quantum Chaos}

Just as in the classical domain there are some investigations on the correlation between regular transport and the onset of quantum chaos, too.
Since the Schr\"odinger equation is a linear equation of motion, the chaos debate in quantum mechanics and the onset of ``quantum chaos'' itself is an open question up to now.
However, Mejia-Monasterio\cite{MejiaMonasterio2005} et al.\ have recently found some evidences of chaotic behavior of a spin chain model under a heat conducting bath coupling.
By passing from ballistic to regular transport as a single system parameter is changed, the system switches from an integrable model into a chaotic one.
This onset of quantum chaos is observed by the change in the level distribution form a Poissonian to a Wigner-Dyson type distribution\cite{Haake2004}.
There have been attempts to find correlations between the change of the level statistic of a system and its transport behavior\cite{Steinigeweg2006I}.

%
%
\section{Description of Nonequilibrium Scenarios}

\subsection{Topological Structure}
\label{chap:OpenSys:sec:1}

Quantum mechanical systems are described by a Hamilton operator $\hamil$ defined in the respective Hilbert space $\Hspace$ of the system.
The class of systems considered here consists of those made up of several subunits, $n$-levels each, coupled by different types of interactions.
Therefore the Hamiltonian of the complete system is described by a local part for each subunit $\mu$, $\hamil_{\ind{loc}}(\mu)$, and an interaction, $\hamil_{\ind{int}}$.
In the following we concentrate on one-dimensional systems (chains) of $N$ subunits and on a next neighbor interaction, only. 
Accordingly, the complete $n^{N}$-dimensional Hamiltonian reads
\begin{equation}
  \label{eq:25}
  \hamil = 
  \sum_{\mu=1}^{N}\hamil_{\ind{loc}}(\mu) + 
  \lambda\sum_{\mu=1}^{N-1}\hamil_{\ind{int}}(\mu,\mu+1)\;,
\end{equation}
with $\lambda$ denoting the coupling strength.
This coupling should be weak in order to guarantee the survival of local features.

A simple example for this model would be a \emph{spin chain}.
In terms of Pauli operators $\hat{\sigma}_i$ ($i=x,y,z$) the local Hamiltonian of a subunit $\mu$ with an energy splitting $\Delta E$ can then be written as
\begin{equation}
  \label{eq:352}
  \hamil_{\ind{loc}}(\mu)
  =\frac{\Delta E}{2}\,\Pauliz(\mu)\;
\end{equation}
while the next neighbor interaction reads
\begin{equation}
  \label{eq:353}
  \hamil_{\ind{int}}(\mu,\mu+1)=
  \sum_{i:x,y,z}c_i
  \hat{\sigma}_i(\mu){\scriptstyle\otimes}\hat{\sigma}_i(\mu+1)\;.
\end{equation}
This is the Heisenberg interaction for $c_x=c_y=c_z=\mbox{const.}$ and the so-called F\"orster coupling for $c_z=0$.

\subsection{The Heat Current}
\label{chap:observables:sec:current}

For a system consisting of several subunits and a Hamiltonian like (\ref{eq:25}), we expect most of the energy to be concentrated in the local part of the Hamiltonian of the system, only a small amount being in the interaction (weak coupling limit).
To get an operator for the current\cite{Lepri2003,Choquard1963,Gemmer2004,Zotos1997,Saito2003} between two adjacent subunits in the system, we consider the time evolution of the local energy operator given by the Heisenberg equation of motion
\begin{equation}
  \label{eq:123}
  \dod{}{t} \hamil_{\ind{loc}}(\mu) 
  = \frac{\I}{\hbar} \Kommutatortxt{\hamil}{\hamil_{\ind{loc}}(\mu)}
  + \pop{\hamil_{\ind{loc}}(\mu)}{t}\;.
\end{equation}
Since here and in the following the local Hamiltonian does not explicitly depend on time, the last term vanishes.
Plugging in the complete Hamiltonian (\ref{eq:25}) and observing the commutator relation we are left with
\begin{equation}
  \label{eq:118}
  \dod{}{t} \hamil_{\ind{loc}}(\mu) 
  = \frac{\I}{\hbar}\lambda\Big(
    \Kommutatortxt{\hamil_{\ind{int}}(\mu-1,\mu)}{\hamil_{\ind{loc}}(\mu)}+
    \Kommutatortxt{\hamil_{\ind{int}}(\mu,\mu+1)}{\hamil_{\ind{loc}}(\mu)}
    \Big)\;.
\end{equation}
In sofar as the local energy is a conserved quantity, this equation may be rewritten as
\begin{equation}
  \label{eq:177}
  \dod{}{t}\hamil_{\ind{loc}}(\mu) 
  = \mbox{div} \hat{J}
  = \hat{J}(\mu,\mu+1)-\hat{J}(\mu-1,\mu)\,,
\end{equation}
where we have introduced the discrete version of the continuity equation.
Thus, the right hand side of (\ref{eq:118}) can be interpreted as a current into as well out of the central subunit $\mu$ and, thus, we define the current operator as 
\begin{equation}
  \label{eq:125}
  \hat{J}(\mu,\mu+1) 
  = \frac{\I}{\hbar}\lambda
    \Kommutatortxt{\hamil_{\ind{int}}(\mu,\mu+1)}{\hamil_{\ind{loc}}(\mu)}\;.
\end{equation}

\subsection{Open Quantum Systems}
\label{chap:OpenSys:sec:2}

The time evolution of the above defined system is governed by the celebrated Liouville-von-Neumann equation
\begin{equation}
  \label{eq:34}
  \dod{\dens}{t} = -\frac{\I}{\hbar}\Kommutatortxt{\hamil}{\dens}\;.
\end{equation}
Switching from the Hilbert space $\Hspace$ to the Liouville space $\Lspace$ it is convenient to write the last equation as
\begin{equation}
  \label{eq:35}
  \dod{}{t}\dens(t) = \opL{L}\,\dens(t)\,,
\end{equation}
defining $\opL{L}$ as a so-called \emph{Liouvillian}\cite{Schack2000}, a super operator acting on operators of the Hilbert space, here the density operator.
Transforming the density operator to a ``vector form'' we can write the Liouvillian as a tensor of second order (matrix) of squared dimension again (in fact a super operator is usually a tensor of higher order). 

The above Liouvillian only describes a coherent time evolution of the closed quantum system.
The von Neumann entropy of the complete system remains constant.
However, the Liouvillian can be extended to a more general form containing also damping processes due to the coupling to an environment.
The environment itself is not described as a concrete physical system, but only by its action on the system.
This coupling leads to equilibration and to a maximum entropy state in the considered system, thus, a thermodynamical situation.

\begin{figure}
  \centering
  \includegraphics[width=11cm]{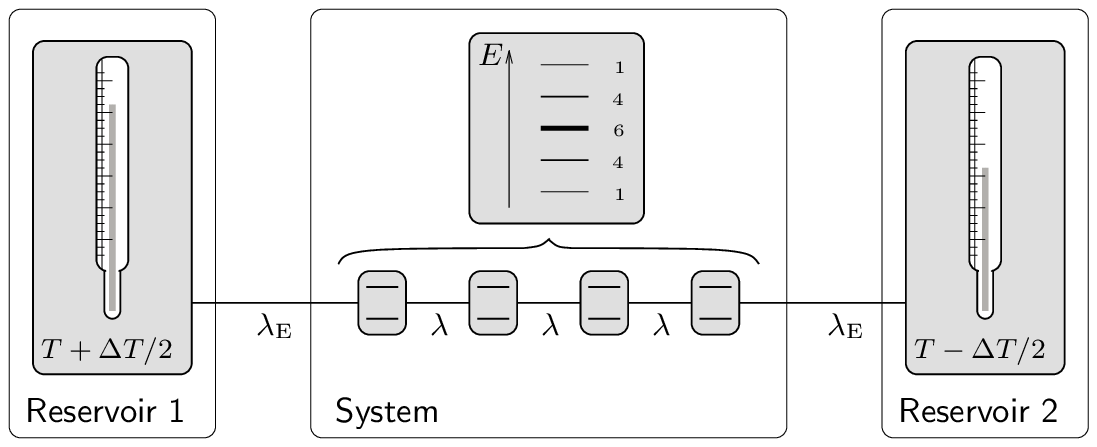}
  \caption{Quantum mechanical heat conduction model system. Several weakly coupled spins ($\lambda$) coupled to two heat reservoirs ($\lambda_{\ind{E}}$) of different temperature modeled by a quantum master equation.}
  \label{fig:2.1}
\end{figure}

The only restriction on the respective super operator $\opL{L}$, whether coherent or not, is that the respective dynamics has to map a density operator again to a density operator.
In 1974 Lindblad\cite{Lindblad1976} was able to introduce the most general (Markovian) form of such an operator.
Thus the time evolution equation for an open quantum system is given by the quantum master equation in Lindblad form\cite{Breuer2002}
\begin{equation}
  \label{eq:112}
  \pop{}{t} \dens
  = \big(\opL{L}_{\ind{S}}+\opL{L}_{\ind{E}}(T,\lambda_{\ind{E}})\big)\;
    \dens\;
\end{equation}
with the coherent part $\opL{L}_{\ind{S}}$ and the dissipator $\opL{L}_{\ind{E}}$. $\lambda_{\ind{E}}$ denotes the system bath coupling strength.
In case of a single two-level system ($n=2$), e.g., controlled by two damping channels with rates $W_{01}$ and $W_{10}$, depending on the bath temperature and the bath coupling strength, we obtain for the dissipator
\begin{eqnarray*}
  \label{eq:72}
  \opL{L}_{\ind{E}}(T,\lambda_{\ind{E}}) \dens
  =&&W_{10}(T,\lambda_{\ind{E}})
     \big(2 \Paulimi \dens \Paulipl
     - \dens \Paulipl\Paulimi - \Paulipl\Paulimi \dens\big)
  \\
  &+& W_{01}(T,\lambda_{\ind{E}})
    \big(2 \Paulipl \dens \Paulimi
     - \dens \Paulimi\Paulipl - \Paulimi\Paulipl \dens\big)\;,
\end{eqnarray*}
in terms of raising and lowering operators $\Paulipl$, $\Paulimi$.
Using this super operator at the edges of a spin chain by simply generalizing the Lindblad operators to raising and lowering operators times the unity of the rest of the system one gets a local bath coupling.
Thus, a heat conduction scenario  is implemented (see \reffig{fig:2.1}).
This procedure has been criticized\cite{Kubo1991}, but for weakly coupled subunits it is a valid method, nevertheless.
A slightly different approach has been used in the literature\cite{Saito2000,Saito2002,Saito2003} by diagonalizing the system first, transforming the operators into this eigenbasis.
However, both approaches lead, at least qualitatively, to the same results.

\subsection{A Heat Conduction Model}
\label{chap:Response:sec:model}

\begin{figure}
\hspace{3mm}
\subfigure[Heisenberg chain]{
  \includegraphics[width=5cm]{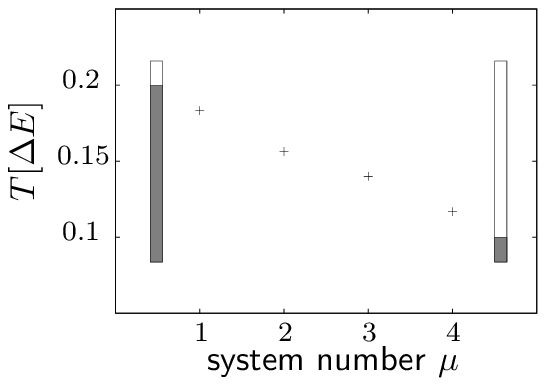}}\quad\quad
\subfigure[F\"orster chain]{
  \includegraphics[width=5cm]{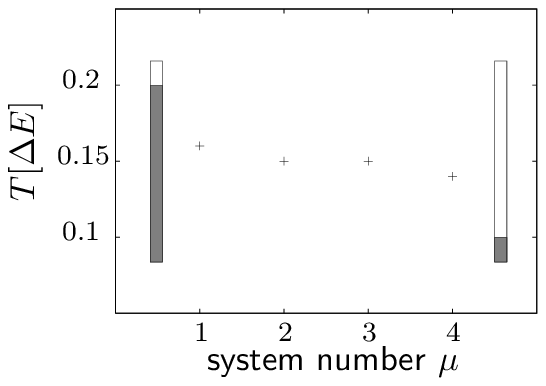}}
\vspace{-3mm}
\caption{Temperature profile for the model in \reffig{fig:2.1} (four spins, bars in the boxes mark the appropriate temperature of the bath). (\textbf{a}) Heisenberg model (\textbf{b}) F\"orster model.}
\label{fig:2.2}
\end{figure}

To describe a proper nonequilibrium scenario we need several ingredients:
a spatially structured system and a local environmental coupling. 
The complete Liouvillian of a heat conduction model system, as depicted in \reffig{fig:2.1}, thus reads
\begin{equation}
  \label{eq:130}
  \opL{L} = \opL{L}_{\ind{S}} 
          + \opL{L}_1(T_1,\lambda_1) + \opL{L}_2(T_2,\lambda_2)\;.
\end{equation}
This operator describes a spatially structured model system (the first term defines the coherent part) coupled to two heat baths of different temperature at either end, given by the dissipators $\opL{L}_1$ and $\opL{L}_2$.

\begin{figure}
\hspace{3mm}
\subfigure[Current]{
  \includegraphics[width=5cm]{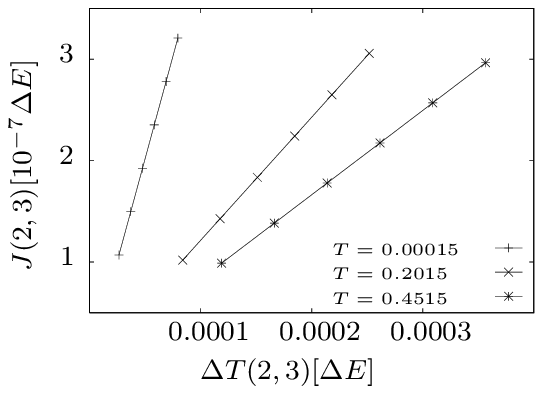}}
\qquad
\subfigure[Conductivity]{
  \includegraphics[width=5cm]{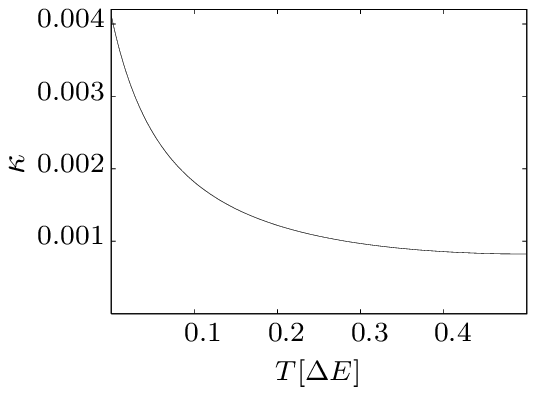}}
\caption{(\textbf{a}) Validity of Fourier's Law for Heisenberg model: Current over temperature difference between the two central systems $(2,3)$. The temperature differences of the heat baths are taken at mean temperatures $T=0.00015,0.2015,0.4515$, respectively. (\textbf{b}) Dependence of the conductivity $\kappa$ on the mean temperature.}
\label{fig:2.3}
\end{figure}

In case of a Heisenberg spin chain, e.g., we find a normal diffusive behavior, i.e. a finite temperature gradient (cf.\ \reffig{fig:2.2}(a)) within the system and Fourier's Law to be fulfilled (cf.\ \reffig{fig:2.3}).
On the other hand, a F\"orster coupled chain shows ballistic behavior, i.e.\ a flat temperature profile and thus a violation of Fourier's Law (see \reffig{fig:2.2}(b)).
See Refs.\cite{Michel2003}, e.g., for a detailed discussion of the transport behavior in open spin chains.

%
%
\section{Perturbation Theory in Liouville Space}

\subsection{Liouville Space Algebra}
\label{chap:Response:sec:super}

Following Schack\cite{Schack2000} et al., the set of linear operators acting on states of an $n$-dimensional Hilbert space $\Hspace$ constitute a $d=n^2$-dimensional complex vector space -- the Liouville space $\Lspace$.
Sorting all entries of an operator $\hat{A}$ in $\Hspace$ into an $n^2$-dimensional vector we can introduce ``ket'' and ``bra'' vectors in $\Lspace$
\begin{equation}
  \label{eq:37}
  \hat{A} \Rightarrow \ketL{\hat{A}}\;,
  \quad
  \hat{A}^{\dagger} \Rightarrow \braL{\hat{A}}\;.
\end{equation}
As a complex vector space one then defines an inner product in $\Lspace$ given by the trace-norm of operators in $\Hspace$
\begin{equation}
  \label{eq:38}
  \sprodL{\hat{A}}{\hat{B}}=\trtxt{\hat{A}^{\dagger}\hat{B}}\;.
\end{equation}
Operators acting on these ``states'' (operators in $\Hspace$) in the Liouville space are called \emph{super operators}.
A super operator $\opL{A}=\ketbraL{\hat{A}}{\hat{B}}$ acts on an arbitrary state $\ketL{\hat{X}}$ of $\Lspace$ according to
\begin{equation}
  \label{eq:39}
  \opL{A}\ketL{\hat{X}} = \ketL{\hat{A}}\sprodL{\hat{B}}{\hat{X}} 
                        = \trtxt{\hat{B}^{\dagger}\hat{X}}\hat{A}\;.
\end{equation}
A complete set of operators $\{\ketL{\hat{A}_j}\}$ in $\Hspace$ constitutes a basis for the Liouville space of the respective system, e.g.\ the set of Pauli operators $\{\hat{1},\Paulix,\Pauliy,\Pauliz\}$ for a spin.
In general this set of operators is complete but not orthonormal, i.e.\ the super operator
\begin{equation}
  \label{eq:40}
  \opL{G} = \sum_j \ketL{\hat{A}_j}\braL{\hat{A}_j}
\end{equation}
is not the unit operator $\opL{1}$ in Liouville space.
According to the definition of the transformation operator in the basis $\{\ketL{\hat{A}_j}\}$ 
\begin{equation}
  \label{eq:41}
  \opL{V}=\Big(\ketL{\hat{A}_1},\cdots,\ketL{\hat{A}_d}\Big)
\end{equation}
one can rewrite (\ref{eq:40}) as
\begin{equation}
  \label{eq:42}
  \opL{G} = \sum_j \ketL{\hat{A}_j}\braL{\hat{A}_j} = \opL{V}\;\opL{V}^{\dagger}\;.
\end{equation}
Thus, the determinant of $\opL{G}$ reads
\begin{equation}
  \label{eq:43}
  \det{\opL{G}} = \det\big(\opL{V}\;\opL{V}^{\dagger}\big)
                = \det{\opL{V}}\det{\opL{V}^{\dagger}}\;.
\end{equation}
All basis operators are linearly independent, $\det{\opL{V}}\neq0$, and thus $\det{\opL{G}}\neq0$, too. 
Therefore the operator $\opL{G}$ is invertible and one finds a dual basis $\ketL{\hat{A}^j}$ according to
\begin{equation}
  \label{eq:44}
  \ketL{\hat{A}^j} = \opL{G}^{-1} \ketL{\hat{A}_j}\;
\end{equation}
with the property
\begin{equation}
  \label{eq:45}
  \sum_j \ketL{\hat{A}^j}\braL{\hat{A}_j} 
  = \sum_j \ketL{\hat{A}_j}\braL{\hat{A}^j} = \opL{1}\;.
\end{equation}
We refer to Refs.\cite{Tarasov2002,Tarasov2002II,Mukamel2003,Caves1999} for a more detailed introduction to super operators.

\subsection{Unperturbed System}
\label{chap:pert:sec:unpert}

The model depicted in \reffig{fig:2.1} is described by the Liouvillian (\ref{eq:130}). 
As the unperturbed system we choose both reservoirs to be at the same temperature $T_1=T_2=T_{\ind{E}}$ with equal coupling strengths $\lambda_1=\lambda_2=\lambda_{\ind{E}}$, defining $\opL{L}_0=\opL{L}_{\ind{S}}+\opL{L}_1(T_{\ind{E}},\lambda_{\ind{E}})+\opL{L}_2(T_{\ind{E}},\lambda_{\ind{E}})$.
Thus, the stationary state of the unperturbed Liouville-von-Neumann equation is a thermal equilibrium state $\hat{\rho}_0$ with temperature $T_{\ind{E}}$:
This state should support neither a heat current nor temperature gradients.

The eigenvalues and eigenvectors of the unperturbed system are given by the eigenequation $\opL{L}_0 \ketL{\hat{\rho}_j} = l_j \ketL{\hat{\rho}_j}$ ($j=0,\dots,n^{2N}-1$).
The above introduced thermal equilibrium state $\hat{\rho}_0$ is an eigenvector of the system with eigenvalue zero, $\opL{L}_0 \ketL{\hat{\rho}_0}=0$ ($l_0=0$).
Since $\hat{\rho}_0$ is the unique stationary state all other eigenvalues have a negative real part.
Furthermore, all eigenvectors except $\hat{\rho}_0$ are trace free\cite{Michel2004}.

\subsection{Local Equilibrium State}
\label{chap:pert:sec:loceq}

The system is perturbed now by applying a small temperature gradient $\Delta T$, defined by the super operator
\begin{equation}
  \label{eq:139}
  \opL{L'}(\Delta T, t) 
  = \opL{L}_1\Big(T+\frac{\Delta T}{2}f(t)\Big) 
  + \opL{L}_2\Big(T-\frac{\Delta T}{2}f(t)\Big)\;,
\end{equation}
with the ``switch-on function''
\begin{equation}
  \label{eq:140}
  f(t) = \Theta(-t)\E^{t}+\Theta(t)
\end{equation}
where $\Theta(t)$ denotes the step function.
Thus, we start at time $t=-\infty$ and allow the perturbation to increase exponentially till $t=0$, setting $\Delta T$ constant for all times $t>0$.
Thereafter the system is subject to this small constant external temperature gradient.
Finally, we are interested in the properties of the stationary \emph{local} equilibrium state of the system reached in the limit $t\rightarrow\infty$.

The time evolution of the whole system under the influence of the perturbation is given by the Liouville-von-Neumann equation
\begin{equation}
  \label{eq:141}
  \pop{}{t}\,\ketL{\hat{\rho}} 
  = \big(\opL{L}_0 + \opL{L'}(\Delta T, t)\big)\,\ketL{\hat{\rho}}\;.
\end{equation}
Starting from a thermal equilibrium state $\hat{\rho}_0$ in the past, we assume the time dependent state of the whole system to be 
\begin{equation}
  \label{eq:142}
  \ketL{\hat{\rho}(t)} 
  = \ketL{\hat{\rho}_0} + \ketL{\Delta\hat{\rho}(t)}\;.
\end{equation}
Introducing this into (\ref{eq:141}), suppressing terms of higher order in the perturbation and observing that $\poptxt{\hat{\rho}_0}{t}=\opL{L}_0\ketL{\hat{\rho}_0}=0$, one finds
\begin{equation}
  \label{eq:143}
  \pop{}{t}\,\ketL{\Delta\hat{\rho}(t)} - \opL{L}_0\,\ketL{\Delta\hat{\rho}(t)}
  = \opL{L'}(\Delta T, t)\,\ketL{\hat{\rho}_0}\;.
\end{equation}
Using the operator transformation in Liouville space as introduced by Kubo\cite{Kubo1991}
\begin{equation}
  \label{eq:144}
  \E^{\opL{L}_0 t} 
  \left(\pop{}{t}\Big[\E^{-\opL{L}_0 t}\ketL{\Delta\hat{\rho}(t)}\Big] \right)
  = \pop{}{t}\ketL{\Delta\hat{\rho}(t)} 
  - \opL{L}_0 \ketL{\Delta\hat{\rho}(t)}
\end{equation}
one finds
\begin{equation}
  \label{eq:357}
  \pop{}{t}\Big[\E^{-\opL{L}_0 t}\ketL{\Delta\hat{\rho}(t)}\Big]
  = \E^{-\opL{L}_0 t}\opL{L'}(\Delta T, t)\,\ketL{\hat{\rho}_0}\;.
\end{equation}
Formally integrating this differential equation, we get
\begin{equation}
  \label{eq:145}
  \ketL{\Delta\hat{\rho}(t)} 
  = \int_{-\infty}^{t} \D t'\,\E^{\opL{L}_0 (t-t')}\,
    \opL{L'}(\Delta T, t')\,\ketL{\hat{\rho}_0}\;.
\end{equation}

\subsection{Standard Kubo-Formula}
\label{chap:pert:sec:kubo}

For the special system described by the Liouvillian
\begin{equation}
  \label{eq:161}
  \pop{\dens}{t} 
  = (\opL{L}_0 + \opL{L'}) \dens  
  = -\frac{\I}{\hbar}\Kommutatortxt{\hamil_{\ind{S}}+\hamil'_t}{\dens}\;,
\end{equation}
one could derive the standard Kubo-formula from (\ref{eq:145}), where the perturbation is considered a potential part $\hamil'_t$ in the system Hamiltonian $\hat{H}_{\ind{S}}$.
Here, we allow for explicitly time-dependent perturbations $\hamil'_t = F(t) \hamil'$ with an explicitly time-independent operator $\hamil'$ and an external field $F(t)$, containing switch on functions like (\ref{eq:140}) as well as some oscillating terms with frequency $\omega$.
Note that also the unperturbed system does no longer contain any thermal reservoir.

The interesting quantity in context of nonequilibrium states is the heat current.
Due to the definition of the current operator in \refsec{chap:observables:sec:current} and the fact that $\dens_0$ does not support any currents one gets
\begin{equation}
  \label{eq:78}
  J = \trtxt{\hat{J}\dens(t)}
    = \trtxt{\hat{J}\dens_0} + \trtxt{\hat{J}\Delta\dens(t)}
    = \trtxt{\hat{J}\Delta\dens(t)}\;.
\end{equation}

Finally, considering the special system (\ref{eq:161}) one arrives at the standard Kubo formula\cite{Michel2006III}
\begin{equation}
  \label{eq:77}
  J
  = \int_{0}^{\infty} \D t'\,F(t'-t)
    \int_{0}^{\beta} \D\beta'\,
    \trtxt{\hat{J}(-t'-\I\hbar\beta')\,\hat{J}\,\hat{\rho}_0}
    \;,   
\end{equation}
where the first current operator refers to the Heisenberg picture (cf.\ (\ref{eq:23})).
Basically, this is a current-current autocorrelation function for the linear response of a quantum system to an external perturbation.

The above method to investigate the perturbation of a quantum system is restricted to perturbations which are defined as a term in the Hamiltonian of the system.
This is \emph{not} the case for thermal perturbations.
These are defined in the Liouville space rather than in the Hilbert space.
Furthermore, it remains unclear how the system reaches the thermal equilibrium state $\hat{\rho}_0$ without any heat bath present.

\subsection{Kubo-Formula in Liouville Space}
\label{chap:pert:sec:liouvillekubo}

We return to the perturbation superoperator (\ref{eq:139}).
This time-dependent superoperator\cite{Michel2004,Michel2006III} could be written as
\begin{equation}
  \label{eq:148}
  \opL{L'}(\Delta T,t) 
  = \frac{\Delta T \lambda_{\ind{E}}}{2} f(t) \opL{E}\;
\end{equation}
with a time independent perturbation operator $\opL{E}$ and a time dependent prefactor.
Thus, (\ref{eq:145}) reduces to
\begin{equation}
  \label{eq:149}
  \ketL{\Delta\hat{\rho}(t)} 
  = \frac{\Delta T\lambda_{\ind{E}}}{2}
    \int_{-\infty}^{t} \D t'\, \E^{\opL{L}_0 (t-t')}
    f(t') \,\opL{E}\, \ketL{\hat{\rho}_0}\;.
\end{equation}
Introducing the unit operator (\ref{eq:45}) in the eigenbasis of the unperturbed system yields
\begin{equation}
  \label{eq:150}
  \ketL{\Delta\hat{\rho}(t)} 
  = \frac{\Delta T\lambda_{\ind{E}}}{2}
    \int_{-\infty}^{t} \D t'\, \E^{\opL{L}_0 (t-t')}
    \sum_j \ketL{\hat{\rho}_j}\braL{\hat{\rho}^j}
    f(t')\, \opL{E}\, \ketL{\hat{\rho}_0}\,.
\end{equation}
With $\E^{\opL{L}_0 (t-t')}\ketL{\hat{\rho}_j}=\E^{l_j(t-t')}\ketL{\hat{\rho}_j}$ we find
\begin{equation}
  \label{eq:151}
  \ketL{\Delta\hat{\rho}(t)} 
  = \frac{\Delta T\lambda_{\ind{E}}}{2}
    \sum_j \braL{\hat{\rho}^j} \opL{E} \ketL{\hat{\rho}_0}\,
    \ketL{\hat{\rho}_j}
    \int_{-\infty}^{t} \E^{l_j(t-t')} f(t')\, \D t'\;.
\end{equation}
By integrating over $t'$ this reduces to
\begin{equation}
  \label{eq:154}
  \ketL{\Delta\hat{\rho}(t)} 
  =\frac{\Delta T\lambda_{\ind{E}}}{2}
   \sum_{j}
   \Big(\frac{\E^{l_j t}}{1-l_j}+\frac{\E^{l_j t}-1}{l_j}\Big)
   \braL{\hat{\rho}^j} \opL{E} \ketL{\hat{\rho}_0}
   \,\ketL{\hat{\rho}_j}\;.
\end{equation}
This perturbative term will include all currents and local temperature gradients of the system under the given perturbation.

Since we are interested in a local equilibrium state -- a stationary state with a constant current and temperature profile, which will be reached after a certain relaxation time -- we consider (\ref{eq:154}) in the limit of $t\rightarrow\infty$ finding
\begin{equation}
  \label{eq:155}
  \ketL{\Delta\hat{\rho}} 
  = \lim_{t\rightarrow\infty} \ketL{\Delta\hat{\rho}(t)}
  = -\frac{\Delta T\lambda_{\ind{E}}}{2}
    \sum_j
    \frac{\braL{\hat{\rho}^j} \opL{E} \ketL{\hat{\rho}_0}}{l_j}\,
    \ketL{\hat{\rho}_j}\;.
\end{equation}
This is the first-order change of the density operator introduced by the perturbation.
Let us call this equation the Kubo-formula in Liouville space\cite{Kubo1991}.
Note the similarity with the Hilbert space perturbation theory: the change of the density operator due to the perturbation depends on both the matrix element of the perturbation operator and all eigenstates of the unperturbed system. 

\subsection{Heat Transport Coefficient}
\label{chap:pert:sec:coeff}

Now we are prepared to account in such Kubo-Liouville scenarios for the local temperature profile and the expectation value of the current.
The stationary density operator of the system is given by $\ketL{\hat{\rho}}=\ketL{\hat{\rho}_0}+\ketL{\Delta\hat{\rho}}$.
Since we know that $\ketL{\hat{\rho}_0}$ does not contribute to any local temperature difference or heat current, the respective expectation value is determined by $\ketL{\Delta\hat{\rho}}$ only.
The local energy difference is given by the operator $\Delta\hat{H}(\mu,\mu+1)=\hat{H}_{\ind{loc}}(\mu)-\hat{H}_{\ind{loc}}(\mu+1)$, and therefore we find for the local internal temperature gradient 
\begin{equation}
  \label{eq:156}
  \delta T(\mu,\mu+1)
  = -\frac{\Delta T\lambda_{\ind{E}}}{2} \sum_j
     \frac{\braL{\hat{\rho}^j} \opL{E} \ketL{\hat{\rho}_0}}{l_j}\,
     \trtxt{\Delta\hat{H}_{\ind{loc}}(\mu,\mu+1)\hat{\rho}_j}
\end{equation}
and the local current within the system
\begin{equation}
  \label{eq:157}
  J(\mu,\mu+1) 
  = -\frac{\Delta T\lambda_{\ind{E}}}{2} \sum_j
     \frac{\braL{\hat{\rho}^j} \opL{E} \ketL{\hat{\rho}_0}}{l_j}\,
     \trtxt{\hat{J}(\mu,\mu+1)\hat{\rho}_j}\;.
\end{equation}
The current as well as the local temperature gradient are thus found to depend linearly on the global temperature difference of the bath systems.
Under stationary conditions the current must be independent of $\mu$, $J(\mu,\mu+1)=J$, so that (\ref{eq:157}) can be rewritten as
\begin{equation}
  \label{eq:158}
  J = - \kappa' \Delta T\;,
\end{equation}
defining the so-called \emph{global conductivity} as
\begin{equation}
  \label{eq:82}
  \kappa' 
  = \frac{\lambda_E}{2}
    \sum_{j=1}^{d-1}
    \frac{\braL{\hat{\rho}^j} \opL{E} \ketL{\hat{\rho}_0}}{l_j}\,
    \trtxt{\hat{J}(\mu,\mu+1)\hat{\rho}_j}\;.
\end{equation}
Eigenstates and eigenvalues entering this global conductivity $\kappa'$ only depend on the mean temperature of the unperturbed system, but not on $\Delta T$. 
Based on this $\kappa'$ as a global property of the system including its contact properties, we call (\ref{eq:158}) \emph{external Fourier's Law}, for which the global conductivity (\ref{eq:82}) defines the overall resistance of our given open quantum system.%

Furthermore, combining (\ref{eq:156}) and (\ref{eq:157}), we define a \emph{local} conductivity as
\begin{equation}
  \label{eq:159}
  \kappa(\mu,\mu+1) 
  = -\frac{J(\mu,\mu+1)}{\delta T(\mu,\mu+1)}
  = -\frac{J}{\delta T(\mu,\mu+1)}
\end{equation}
implying $\kappa(\mu,\mu+1)$ to be independent of the external difference $\Delta T$.

This new approach does not have the problem of introducing a potential term into the Hamiltonian of the system, like in standard Kubo formulas for heat conduction.
The bath systems, modeled by a quantum master equation, directly define the perturbation in Liouville space.
Like in standard perturbation theory in Hilbert space the first order correction to the stationary state of the system is expressed in terms of transition matrix elements of the perturbation operator and the eigenstates and eigenvalues of the unperturbed system.
Only the non-orthogonality of the eigensystem of the unperturbed Liouvillian calls for a more careful treatment; formally the equations are very similar.

%
%
\section{Hilbert Space Average Method for Heat Conduction}

As noted already by Einstein\cite{Einstein1905,Einstein1908} in his study of Brownian motion, there is a close connection between the bath assisted transport as considered in the last Section and the decay from a nonequilibrium initial state into the global equilibrium within a closed system.
An important role should be played by the transport coefficients as material constants.
However, in small closed quantum systems we do not find a regular decay behavior, but rather coherent oscillations.
Unfortunately, there are neither theoretical nor numerical investigations of larger systems feasible at the moment.
Thus, we turn to yet another approach to heat transport within quantum mechanics based on the Hilbert Space Average Method (HAM\cite{Gemmer2003,Gemmer2004,Gemmer2005I,Michel2006I,Michel2005II}).
This method allows for a prediction of the Schr\"odinger evolution of certain ``coarse grained'' observables, such as, e.g.\ the occupation probability of a whole energy band, and is thus well suited for the investigation of the decay in mesoscopic quantum systems.

\subsection{Mesoscopic Model System}

\begin{figure}
  \centering 
  \includegraphics[width=7.2cm]{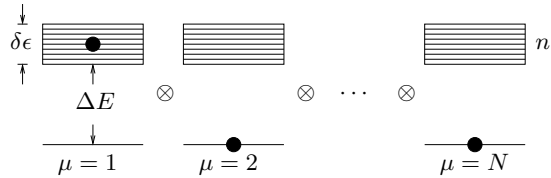}
  \caption{Heat conduction model: $N$ coupled subunits with ground level and band of $n$ equally distributed levels. Black dots refer to used initial states.}
\label{fig:4.1}  
\end{figure}

The class of systems we are going to analyze next is depicted in \reffig{fig:4.1}, again characterized by the Hamiltonian (\ref{eq:25}). 
Instead of using two level systems the $N$ identical subunits now feature a non-degenerate ground state and a band of $n$ excited states each, equally distributed over some band width $\delta\epsilon$ in such a way that the band width is small compared to the local energy gap $\Delta E$ ($\delta\epsilon\ll\Delta E$, $\delta\epsilon$ in units of $\Delta E$).
These subunits are coupled by an energy exchanging next neighbor interaction $\hat{H}_{\ind{int}}(\mu,\mu+1)$, chosen to be a (normalized) random Hermitian matrix allowing for any possible transition such as to avoid any bias. 
Our results will turn out to be independent of the exact form of the matrix.
We choose the next neighbor coupling to be weak compared to the local gap ($\lambda\ll\Delta E$, $\lambda$ in units of $\Delta E$). 
This way the full energy is approximately given by the sum of the local energies and these are approximately given by $\bra{\psi(t)}\hat{H}_{\ind{loc}}(\mu)\ket{\psi(t)}$, where $P_{\mu}$ is the probability to find the $\mu$-th subsystem in its excited state.
This model is primarily meant to demonstrate how energy transport might emerge from Schr\"odinger dynamics -- a statistical diffusive behavior from a time reversible microscopic theory.

\begin{figure}
  \centering 
  \includegraphics[width=6.6cm]{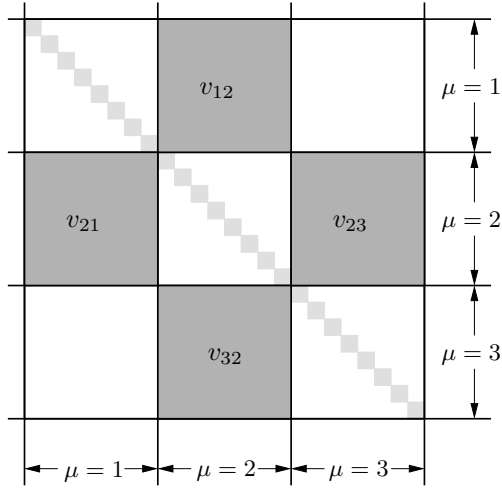}
  \caption{Reduced Hamiltonian (local part light gray, interaction dark gray) of the one excitation subspace in order to investigate heat conduction in the model displayed in \reffig{fig:4.1} ($N=3$ subunits).}
\label{fig:4.2}  
\end{figure}

Restricting ourselves to the single excitation subspace, i.e.\ initial states with one system somewhere in the upper band, all others in their ground state, we can derive a reduced Hamiltonian model.
The respective Hamiltonian matrix is shown in \reffig{fig:4.2}, here for three subsystems only.
The dark gray blocks refer to the interaction $v_{ij}$ between adjacent subunits, whereas the light gray ones to the local Hamiltonian.
We skip all diagonal couplings, transitions within the band, for simplicity.

We introduce projection operators $\hat{P}_{\mu}$.
These operators are matrices with the unit operator of dimension $n$ at position $(\mu,\mu)$ and zeros elsewhere.
Thus, the operator $\hat{P}_{\mu}$ projects out the part of the state describing the $\mu$th system being in the upper band.
The probability of finding subunit $\mu$ excited is thus given by $P_{\mu}=\text{Tr}\{\ket{\psi}\bra{\psi}\hat{P}_{\mu}\}$.

\section{Hilbert Space Average Method}
\label{chap:hamtr:sec:dyson}

The Hilbert space average method (HAM) is in essence a technique to guess the value of some quantities defined as a function of $\ket{\psi}$ if $\ket{\psi}$ itself is not known in full detail, only some features of it. 
Here the method is used to guess the expectation value $\bra{\psi}\hat{A}\ket{\psi}$ if the only information about $\ket{\psi}$ is the set of expectation values $\bra{\psi}\hat{P}_{\mu}\ket{\psi}=P_{\mu}$. 
Such a statement naturally has to be a guess, since there are in general many different $\ket{\psi}$ that are in accord with the given set of $P_{\mu}$, but produce different values for $\bra{\psi}\hat{A}\ket{\psi}$. 
Thus, the key question for the reliability of this guess is whether the distribution of the $\bra{\psi}\hat{A}\ket{\psi}$ produced by the respective set of $\ket{\psi}$'s is broad or whether almost all those $\ket{\psi}$ yield $\bra{\psi}\hat{A}\ket{\psi}$ that are approximately equal. 
It can be shown that if the spectral width of $\hat{A}$ is not too large and  $\hat{A}$ is high-dimensional almost all individual $\ket{\psi}$ yield an expectation value close to the mean of the distribution of the $\bra{\psi}\hat{A}\ket{\psi}$ and thus the HAM guess will be reliable\cite{Gemmer2004}.
To find that mean one has to average with respect to the  $\ket{\psi}$. 
This is called a Hilbert space average $A$ and will be denoted as 
\begin{equation}
  \label{eq:4}
  A=\trtxt{\hat{A}\hat{\alpha}}  
  \quad \text{with}\quad 
  \hat{\alpha}
  :=\Haver{\ket{\psi}\bra{\psi}}_{\{\bra{\psi}\hat{P}_{\mu}\ket{\psi}=P_{\mu}\}}\;.
\end{equation}
This expression stands for the average of $\bra{\psi}\hat{A}\ket{\psi}=\trtxt{\hat{A}\ket{\psi}\bra{\psi}}$ over all $\ket{\psi}$ that feature $\bra{\psi}\hat{P}_{\mu}\ket{\psi}=P_{\mu}$ but are uniformly distributed otherwise. 
Uniform distribution implies invariance with respect to all unitary transformations $\hat{U}$ that leave $P_{\mu}$ unchanged, i.e., $\bra{\psi}\hat{U}^{\dagger}\hat{P}_{\mu}\hat{U}\ket{\psi}=P_{\mu}$.
Thus the $\hat{U}$ are specified by $[\hat{U},\hat{P}_{\mu}]=0$.

Any such $\hat{U}$ has to leave $\hat{\alpha}$ invariant, i.e., $[\hat{U},\hat{\alpha}]=0$. 
Furthermore, $\hat{\alpha}$ has to obey $\trtxt{\hat{\alpha}\hat{P}_{\mu}}=P_{\mu}$. 
These conditions\cite{Michel2005,Michel2006III} uniquely determine $\hat{\alpha}$ as
\begin{equation}
  \label{eq:5}
  \hat{\alpha}=\sum_{\mu} \frac{P_{\mu}}{n} \,\hat{P}_{\mu}\;,
\end{equation}
i.e., an expansion in terms of the projectors $\hat{P}_{\mu}$ themselves.

Consider a pure state of the total system at some time $t$, $\ket{\psi(t)}$. 
In the interaction picture the dynamics of the full system is controlled by the interaction $\hat{V}(t)$, only. 
The time evolution is generated by the corresponding Dyson series truncated to second order in a weak interaction scenario  
\begin{equation}
  \label{eq:222}
  \ket{\psi(t+\tau)}
  \approx\big(\hat{1}-\frac{\I}{\hbar}\hat{U}_{1}(\tau)
                     -\frac{1}{\hbar^2}\hat{U}_{2}(\tau)\big)\ket{\psi(t)}
  = \hat{D}(\tau)\ket{\psi(t)}\;,
\end{equation}
with the two time evolution operators
\begin{equation}
  \label{eq:229}
  \hat{U}_{1}(\tau) = \int_0^{\tau}\D\tau'\,\hat{V}(\tau')\;, \quad 
  \hat{U}_{2}(\tau) = \int_0^{\tau}\D\tau'\int_0^{\tau'}\D\tau''
                    \,\hat{V}(\tau') \,\hat{V}(\tau'')
\end{equation}
and the interaction Hamiltonian $\hat{V}(t)$ in the interaction picture.
This allows for the computation of the probabilities $P_{\mu}$ at time $t+\tau$ finding the $\mu$th system somewhere in its excited band
\begin{equation}
  \label{eq:6}
  P_{\mu}(t+\tau)=\trtxt{\hat{D}(\tau)\,\ket{\psi(t)}\bra{\psi(t)}\,
                         \hat{D}^{\dagger}(\tau)\,\hat{P}_{\mu}}\;.
\end{equation}
Assume that rather than $\ket{\psi(t)}$ itself only the set of expectation values $P_{\mu} $ is known. 
The application of HAM produces a guess for the $P_{\mu} (t+\tau)$ based on the  $P_{\mu}(t)$ replacing $\ket{\psi(t)}\bra{\psi(t)}$ by the above $\hat{\alpha}$ (\ref{eq:5})
\begin{equation}
  \label{eq:7}
  P_{\mu}(t+\tau)
  \approx
  \trtxt{\hat{D}(\tau)\,\hat{\alpha}\,\hat{D}^{\dagger}(\tau)\,\hat{P_{\mu}}}
  \approx 
  \sum_{\nu}\frac{P_{\nu}}{n}\,
  \trtxt{\hat{D}(\tau)\,\hat{P_{\nu}}\,\hat{D}^{\dagger}(\tau)\,\hat{P_{\mu}}}
  \;.
\end{equation}
Using (\ref{eq:222}) this yields after lengthy, but rather straight forward calculations 
\begin{equation}
  \label{eq:8} 
  P_{\mu}(t+\tau)-P_{\mu}(t)
  \approx f(\tau)\,\big(P_{\mu-1}(t)+P_{\mu+1}(t)-2P_{\mu}(t)\big)\;,
\end{equation}
with
\begin{equation}
  \label{eq:9}
  f(\tau)
  := \frac{2}{n}
      \int_0^{\tau}\int_0^{\tau'}
      \trtxt{\hat{V}(\tau'')\,\hat{V}(0)}
      \,\D\tau''\,\D\tau'\;.
\end{equation}
The above integrand  is essentially  the same environmental correlation function that appears in the memory kernels of standard projection operator techniques.
Those correlation functions typically feature some decay time $\tau_c$ after which they vanish. 
Integrating twice yields functions that increase linear in time after $\tau_c$. 
Hence, for $\tau>\tau_c$ one simply gets $f(\tau)=\gamma\tau$, where $\gamma$ has to be computed from (\ref{eq:9}), but typically corresponds to a transition rate as obtained from Fermi's Golden Rule, i.e.,
\begin{equation}
  \label{eq:10}
  \gamma = \frac{2\pi\lambda^2 n}{\hbar\delta\epsilon}\;.
\end{equation} 
Assuming that the decay times of the correlation functions are small one can transform the iteration scheme (\ref{eq:8}) into a set of differential equations
\begin{eqnarray}
  \label{eq:4a}
  \dod{P_1}{t} &=& - \gamma(P_1 - P_2)\;, \\
  \label{eq:4b}
  \dod{P_{\mu}}{t} &=& - \gamma (2 P_{\mu} - P_{\mu-1} - P_{\mu+1})\;,\\ 
  \label{eq:4c}
  \dod{P_N}{t} &=& - \gamma (P_N - P_{N-1})\;.
\end{eqnarray}

The above scheme, however, only applies if the dynamics of the system are reasonably well described by a Dyson series truncated at second order also for times $\tau$ larger than $\tau_c$. 
This and similar arguments yield the following necessary conditions for the above described occurrence of diffusive transport
\begin{equation}
  \label{eq:11}
  2 \lambda\frac{n}{\delta\epsilon} \geq 1, 
  \quad \lambda^2\frac{n}{\delta\epsilon^2} \ll 1\;.
\end{equation}
If those criteria are violated no diffusive transport is expected.
For a more detailed description of HAM and its various implications, see Gemmer\cite{Gemmer2005I,Gemmer2006I,Gemmer2004}.

\subsection{Complete Solution}

\begin{figure}
  \hspace{3mm}
  \subfigure[Decay]{
    \includegraphics[width=5cm]{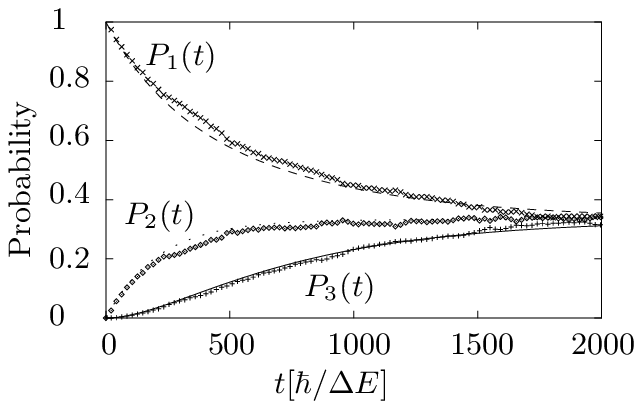}}
  \qquad
  \subfigure[Accuracy]{
    \includegraphics[width=5cm]{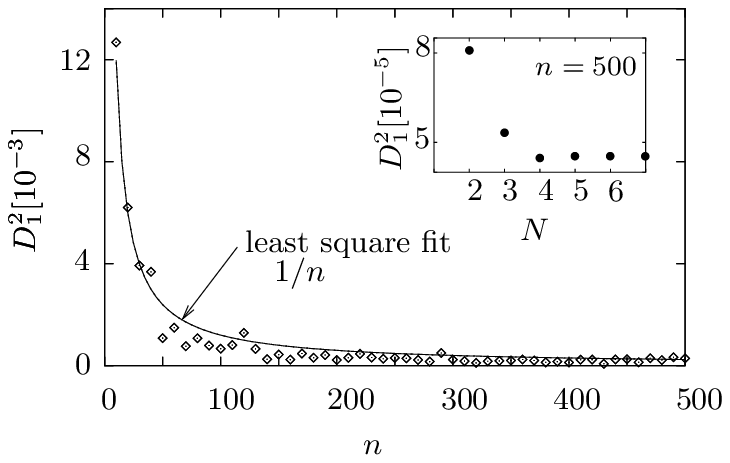}}
  \caption{(\textbf{a}) Probability to find the excitation in the $\mu=1,2,3$ system. Comparison of the HAM prediction (lines) and the exact Schr\"odinger solution (dots). ($N=3$, $n=500$, $\lambda=5\cdot 10^{-5}$, $\delta\epsilon=0.05$) (\textbf{b}) Deviation of HAM from the exact solution ($D^2_1$ least squares): dependence of $D^2_1$ on $n$ for $N=3$. Inset: dependence of $D^2_1$ on $N$ for $n=500$.} 
\label{fig:4.3}
\end{figure}

To analyze validity and performance of HAM we compare its results  with data from a direct numerical integration of the Schr\"odinger equation. 
This is, of course, only possible for systems small enough to allow for the latter.  
Hereby we restrict ourselves to initial states with only one subsystem in the exited band (all others in the ground level, black dots in Fig.~\ref{fig:4.1}). 
Finding an effective Hamiltonian for the one-excitation subspace we are able to solve the Schr\"odinger equation for up to $N=10$ subsystems, $n=500$ levels each.
Firstly restricting to $N=3$, the numerical results together with the HAM predictions for an initial state with $P_1=1$, $P_2=P_3=0$ are shown in \reffig{fig:4.3}(a) ($N=3$, $n=500$, $\delta\epsilon = 0.05$, $\lambda=5\cdot10^{-5}$). 
There is a good agreement.

To investigate the accuracy of the HAM for, e.g., $P_1(t)$ we introduce $D^2_1$, being the time-averaged quadratic deviation of the exact (Schr\"odinger) result for $P_1(t)$ from the HAM prediction
\begin{equation}  
  \label{eq:12}
  D^2_1=\frac{\gamma}{5}
  \int_0^{\frac{5}{\gamma}}(P_1^{\ind{HAM}}(t)-P_1^{\ind{exact}}(t))^2\D t\;.
\end{equation}
To analyze how big this ``typical deviation'' is, we have computed $D^2_1$ for a $N=3$ system with the first subunit initially in an arbitrary excited state, for different numbers of states $n$ in the bands. 
As shown in \reffig{fig:4.3}(b), the deviation scales like $1/n$ with the band size, i.e., vanishes in the limit of high dimensional subunits. 
This behavior does not come as a complete surprise since it has theoretically been conjectured and numerically verified in the context of equilibrium fluctuations of non-Markovian systems\cite{Gemmer2004,BorowskiGemmer2003}. 
The inset of \reffig{fig:4.3}(b) shows that $D_1^2$ goes down also with increasing number of subunits $N$ but then levels off. 
Altogether HAM appears to be applicable even down to moderately sized systems. 
So far we have restricted ourselves to pure states. 
A drastic further reduction of $D_1^2$ can be expected for mixed states (which are typical in the context of thermodynamical phenomena), since pure state fluctuations should cancel partially if added together.

\subsection{Energy Transport}
\label{chap:trenscoeff:sec:energy}

In order to account for the energy diffusion constant in the given model system, we have to consider the energy current inside the system.
The total internal energy in one subunit of the above described type is given by the probability to be in the excited band of the subunit times the width of the energy gap, $U_{\mu}=\Delta E P_{\mu}$.
Thus the current is defined by the change of the internal energy $U_{\mu}$ of two adjacent subunits
\begin{equation}
  \label{eq:285}
  J 
  = \frac{1}{2}\left(\dod{U_{1}}{t}-\dod{U_{2}}{t}\right)
  = \frac{\Delta E}{2}\left(\dod{P_{1}}{t}-\dod{P_{2}}{t}\right)\;.
\end{equation}
The change of the probability in time is given by the rate equation (\ref{eq:4a}) and one finds the current
\begin{equation}
  \label{eq:287}
  J 
  = -\kappa\Delta E\big(P_2-P_1\big)=-\kappa(U_2-U_1)\;.
\end{equation}
Thus, the current is a linear function of the probability gradient respectively \emph{energy gradient} inside the system.
This specifies a diffusive behavior since energy spreads through the system in a statistical way.
Therefore we may extract the energy diffusion constant from the above equation, identifying the parameter $\kappa$ from the rate equation with the transport coefficient given in (\ref{eq:10}).
This energy diffusion constant depends on the coupling strength of the subunits $\lambda$ and on the state density of the excited band $n/\delta\epsilon$.
Remarkably this diffusive behavior is not restricted to an initially small energy gradient. 

\subsection{Heat Transport}
\label{chap:trenscoeff:sec:heat}

\begin{figure}
  \centering
  \includegraphics[width=6cm]{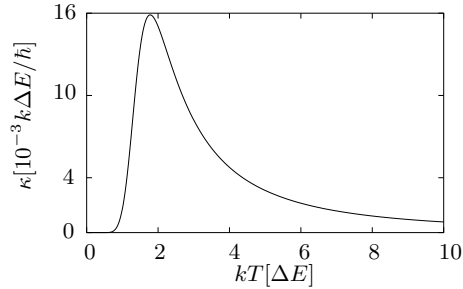}
  \caption{Heat conductivity (\ref{eq:13}) over temperature for a system with $n=500$, $\delta\epsilon = 0.05$ and $\lambda=5\cdot10^{-5}$.}
\label{fig:4.4}  
\end{figure}

So far we have considered energy diffusion through the system only.
The final state should approach equipartition of energy over all subunits -- a thermal equilibrium state\cite{Gemmer2004}.
Close to this equilibrium we expect the system to be in a state where the probability distribution of each subunit is approximately canonical (Gibbs state), but still with a slightly different temperature $T_{\mu}$ for each site.
Specializing in those ``local equilibrium states'' and exploiting the HAM results allows for a direct connection of the local energy current between any two adjacent subunits with their temperature gradient $\Delta T=T_1-T_2$ and their mean temperature $T=(T_1+T_2)/2$. 
Since this connection is found to be linear in the temperature gradient one can simply read off the temperature dependent \emph{heat conductivity}\cite{Michel2005} 
\begin{equation}
  \label{eq:13}
  \kappa = \frac{2\pi k\lambda ^2 n^2}{\hbar \delta \epsilon} \;
           \left(\frac{\Delta E}{kT}\right)^2\;
           \frac{\E^{-\frac{\Delta E}{kT}}}{\Big(1+n\E^{-\frac{\Delta E}{kT}}\Big)^2}\;,
\end{equation}
as displayed in Fig.~\ref{fig:4.4}. 
For the present model this is in agreement with $\kappa=\gamma c$ according to (\ref{eq:10}), if one inserts for $c$ the specific heat of one subunit.

%
%
\section{Conclusion}

Is the question concerning the microscopic foundation of Fourier's Law completely answered now?
Certainly not in all details, however, there are new building blocks and perspectives accessible so that a satisfying answer in the near future should come into reach.

Both, the Hilbert Space Average Method and the perturbation theory in Liouville space, have opened new perspectives towards the investigation of statistical behavior in small quantum systems near the global equilibrium.
Several interesting open questions remain which hopefully will be answered in the near future.
Those questions mainly address the relationship between the presented new methods, the Hilbert Space Average Method and the Kubo formula in Liouville space as well as the connection to some other important theoretical concepts like the standard Kubo formula, the quasiparticle approach, quantum chaos etc.
Work is in progress to apply the new methods to other concrete models and to discuss differences between the theoretical predictions. 
However, until now the application of the Hilbert Space Average Method on models like spin chains remain challenging.

In face of the rapid miniaturization of many technical devices like e.g.\ computer chips, and the trend to utilize quantum mechanical systems, a deeper understanding of transport and relaxation processes and the thermodynamic properties of small systems will become necessary in the near future.
At least, it will be indispensable to know to what extent the thermodynamic concepts can be extended to the nanoscale and which limits should be taken into account.
Besides the foundation of relaxation and transport on quantum mechanics also new aspects beyond a proper thermodynamical description are interesting.
For example, there are some materials featuring extremely high thermal conductivities, due to magnetic excitations.
Here we found that besides the emergence of equilibrium properties in small quantum systems, nonequilibrium thermodynamic aspects are already present far below the thermodynamic limit.
Hopefully, the ongoing research in this field will help to shed new light on the fundamental aspects behind those interesting and long since known topics of relaxation and transport.



\begin{thebibliography}{91}
\expandafter\ifx\csname natexlab\endcsname\relax\def\natexlab#1{#1}\fi
\expandafter\ifx\csname bibnamefont\endcsname\relax
  \def\bibnamefont#1{#1}\fi
\expandafter\ifx\csname bibfnamefont\endcsname\relax
  \def\bibfnamefont#1{#1}\fi
\expandafter\ifx\csname citenamefont\endcsname\relax
  \def\citenamefont#1{#1}\fi
\expandafter\ifx\csname url\endcsname\relax
  \def\url#1{\texttt{#1}}\fi
\expandafter\ifx\csname urlprefix\endcsname\relax\def\urlprefix{URL }\fi
\providecommand{\bibinfo}[2]{#2}
\providecommand{\eprint}[2][]{\url{#2}}

\bibitem[{\citenamefont{Bonetto et~al.}(2000)\citenamefont{Bonetto, Lebowitz,
  and Rey-Bellet}}]{Bonetto2000}
\bibinfo{author}{\bibfnamefont{F.}~\bibnamefont{Bonetto}},
  \bibinfo{author}{\bibfnamefont{J.}~\bibnamefont{Lebowitz}}, \bibnamefont{and}
  \bibinfo{author}{\bibfnamefont{L.}~\bibnamefont{Rey-Bellet}},
  \emph{\bibinfo{title}{Mathematical Physics 2000}} (\bibinfo{publisher}{World
  Scientific Publishing Company}, \bibinfo{year}{2000}), chap.
  \bibinfo{chapter}{"Fourier's Law: A Challenge to Theorists"}, pp.
  \bibinfo{pages}{128--150}, \bibinfo{note}{also published at
  arXiv:math-ph/0002052}.

\bibitem[{\citenamefont{Lepri et~al.}(2003)\citenamefont{Lepri, Livi, and
  Politi}}]{Lepri2003}
\bibinfo{author}{\bibfnamefont{S.}~\bibnamefont{Lepri}},
  \bibinfo{author}{\bibfnamefont{R.}~\bibnamefont{Livi}}, \bibnamefont{and}
  \bibinfo{author}{\bibfnamefont{A.}~\bibnamefont{Politi}},
  \bibinfo{journal}{Phys. Rep.} \textbf{\bibinfo{volume}{377}},
  \bibinfo{pages}{1} (\bibinfo{year}{2003}).

\bibitem[{\citenamefont{Fourier}(1955)}]{Fourier1955}
\bibinfo{author}{\bibfnamefont{J.}~\bibnamefont{Fourier}},
  \emph{\bibinfo{title}{The analytical theory of heat}}
  (\bibinfo{publisher}{Dover Publ.}, \bibinfo{address}{New York},
  \bibinfo{year}{1955}), \bibinfo{note}{transl. by Alexander Freeman. Original:
  Th\'eorie analytique de la chaleur (publ. 1822). Unabridged reprinting of the
  engl. Transl. anno 1878.}

\bibitem[{\citenamefont{Debye}(1914)}]{Debye1914}
\bibinfo{author}{\bibfnamefont{P.}~\bibnamefont{Debye}}, in
  \emph{\bibinfo{booktitle}{Vortr\"age \"uber die kinetische Theorie der
  Materie und der Elektrizit\"at}}, edited by
  \bibinfo{editor}{\bibfnamefont{M.}~\bibnamefont{Planck}}
  (\bibinfo{publisher}{Teubner}, \bibinfo{address}{Leipzig},
  \bibinfo{year}{1914}), vol.~\bibinfo{volume}{VI} of
  \emph{\bibinfo{series}{Mathematische Vortr\"age an der Universit\"at
  G\"ottingen}}, pp. \bibinfo{pages}{19--60}, \bibinfo{note}{mathematische
  Vorlesungen an der Universit\"at G\"ottingen: VI}.

\bibitem[{\citenamefont{Kittel}(1969)}]{Kittel1969}
\bibinfo{author}{\bibfnamefont{C.}~\bibnamefont{Kittel}},
  \emph{\bibinfo{title}{Einf\"uhrung in die Festk\"orperphysik}}
  (\bibinfo{publisher}{R. Oldenbourg Verlag}, \bibinfo{address}{M\"unchen,
  Wien}, \bibinfo{year}{1969}).

\bibitem[{\citenamefont{Peierls}(1929)}]{Peierls1929}
\bibinfo{author}{\bibfnamefont{R.}~\bibnamefont{Peierls}},
  \bibinfo{journal}{Ann. Phys. (N. Y.)} \textbf{\bibinfo{volume}{3}},
  \bibinfo{pages}{1055} (\bibinfo{year}{1929}).

\bibitem[{\citenamefont{Peierls}(2001)}]{Peierls1955}
\bibinfo{author}{\bibfnamefont{R.}~\bibnamefont{Peierls}},
  \emph{\bibinfo{title}{Quantum Theory of Solids}}
  (\bibinfo{publisher}{Clarendon Press}, \bibinfo{address}{Oxford},
  \bibinfo{year}{2001}).

\bibitem[{\citenamefont{Michel and Wagner}(1976)}]{Michel1976}
\bibinfo{author}{\bibfnamefont{H.}~\bibnamefont{Michel}} \bibnamefont{and}
  \bibinfo{author}{\bibfnamefont{M.}~\bibnamefont{Wagner}},
  \bibinfo{journal}{Phys. Status Solidi B} \textbf{\bibinfo{volume}{75}},
  \bibinfo{pages}{507} (\bibinfo{year}{1976}).

\bibitem[{\citenamefont{Michel and Wagner}(1978{\natexlab{a}})}]{Michel1978}
\bibinfo{author}{\bibfnamefont{H.}~\bibnamefont{Michel}} \bibnamefont{and}
  \bibinfo{author}{\bibfnamefont{M.}~\bibnamefont{Wagner}},
  \bibinfo{journal}{Ann. Phys. (Leipzig)} \textbf{\bibinfo{volume}{35}},
  \bibinfo{pages}{425} (\bibinfo{year}{1978}{\natexlab{a}}).

\bibitem[{\citenamefont{Michel and Wagner}(1978{\natexlab{b}})}]{Michel1978II}
\bibinfo{author}{\bibfnamefont{H.}~\bibnamefont{Michel}} \bibnamefont{and}
  \bibinfo{author}{\bibfnamefont{M.}~\bibnamefont{Wagner}},
  \bibinfo{journal}{Phys. Status Solidi B} \textbf{\bibinfo{volume}{85}},
  \bibinfo{pages}{195} (\bibinfo{year}{1978}{\natexlab{b}}).

\bibitem[{\citenamefont{Ziman}(2001)}]{Ziman2001}
\bibinfo{author}{\bibfnamefont{J.}~\bibnamefont{Ziman}},
  \emph{\bibinfo{title}{Electrons and Phonons}} (\bibinfo{publisher}{Oxford:
  Clarendon Press}, \bibinfo{address}{Oxford}, \bibinfo{year}{2001}).

\bibitem[{\citenamefont{Wagner}(1999)}]{Wagner1999}
\bibinfo{author}{\bibfnamefont{M.}~\bibnamefont{Wagner}},
  \bibinfo{journal}{Phil. Mag. B} \textbf{\bibinfo{volume}{79}},
  \bibinfo{pages}{1839} (\bibinfo{year}{1999}).

\bibitem[{\citenamefont{Fermi et~al.}(1965)\citenamefont{Fermi, Pasta, and
  Ulam}}]{Fermi1955}
\bibinfo{author}{\bibfnamefont{E.}~\bibnamefont{Fermi}},
  \bibinfo{author}{\bibfnamefont{J.}~\bibnamefont{Pasta}}, \bibnamefont{and}
  \bibinfo{author}{\bibfnamefont{S.}~\bibnamefont{Ulam}},
  \emph{\bibinfo{title}{Note e Memorie (Collected Papers)}}
  (\bibinfo{publisher}{The University of Chicago Press},
  \bibinfo{address}{Roma}, \bibinfo{year}{1965}), vol. \bibinfo{volume}{II
  (United States 1939-1954)}, chap. \bibinfo{chapter}{"Studies of non Linear
  Problems"}, pp. \bibinfo{pages}{977--988}.

\bibitem[{\citenamefont{Kubo}(1957)}]{Kubo1957}
\bibinfo{author}{\bibfnamefont{R.}~\bibnamefont{Kubo}}, \bibinfo{journal}{J.
  Phys. Soc. Jpn.} \textbf{\bibinfo{volume}{12}}, \bibinfo{pages}{570}
  (\bibinfo{year}{1957}).

\bibitem[{\citenamefont{Kubo et~al.}(1991)\citenamefont{Kubo, Toda, and
  Hashitsume}}]{Kubo1991}
\bibinfo{author}{\bibfnamefont{R.}~\bibnamefont{Kubo}},
  \bibinfo{author}{\bibfnamefont{M.}~\bibnamefont{Toda}}, \bibnamefont{and}
  \bibinfo{author}{\bibfnamefont{N.}~\bibnamefont{Hashitsume}},
  \emph{\bibinfo{title}{Statistical {P}hysics {II}: {N}onequilibrium
  {S}tatistical {M}echanics}}, no.~\bibinfo{number}{31} in
  \bibinfo{series}{Solid-State Sciences} (\bibinfo{publisher}{Springer},
  \bibinfo{address}{Berlin, Heidelberg, New-York}, \bibinfo{year}{1991}),
  \bibinfo{edition}{2nd} ed.

\bibitem[{\citenamefont{Mori}(1959)}]{Mori1956}
\bibinfo{author}{\bibfnamefont{H.}~\bibnamefont{Mori}}, \bibinfo{journal}{Phys.
  Rev.} \textbf{\bibinfo{volume}{115}}, \bibinfo{pages}{298}
  (\bibinfo{year}{1959}).

\bibitem[{\citenamefont{Mahan}(2000)}]{Mahan1981}
\bibinfo{author}{\bibfnamefont{G.~D.} \bibnamefont{Mahan}},
  \emph{\bibinfo{title}{Many-{P}article {P}hysics}} (\bibinfo{publisher}{Plenum
  Press}, \bibinfo{address}{New York, London}, \bibinfo{year}{2000}),
  \bibinfo{edition}{3rd} ed.

\bibitem[{\citenamefont{de~Groot and Mazur}(1962)}]{Groot1962}
\bibinfo{author}{\bibfnamefont{S.}~\bibnamefont{de~Groot}} \bibnamefont{and}
  \bibinfo{author}{\bibfnamefont{P.}~\bibnamefont{Mazur}},
  \emph{\bibinfo{title}{Non-equilibrium {T}hermodynamics}}
  (\bibinfo{publisher}{North-Holland Publ. Comp.},
  \bibinfo{address}{Amsterdam}, \bibinfo{year}{1962}).

\bibitem[{\citenamefont{Luttinger}(1964)}]{Luttinger1964}
\bibinfo{author}{\bibfnamefont{J.~M.} \bibnamefont{Luttinger}},
  \bibinfo{journal}{Phys. Rev.} \textbf{\bibinfo{volume}{135}},
  \bibinfo{pages}{A1505} (\bibinfo{year}{1964}).

\bibitem[{\citenamefont{Heidrich-Meisner}(2005)}]{Heidrich2005}
\bibinfo{author}{\bibfnamefont{F.}~\bibnamefont{Heidrich-Meisner}},
  \bibinfo{type}{Dissertation}, \bibinfo{school}{Technische Universit\"at
  Braunschweig} (\bibinfo{year}{2005}).

\bibitem[{\citenamefont{Zotos et~al.}(1997)\citenamefont{Zotos, Naef, and
  Prelovsek}}]{Zotos1997}
\bibinfo{author}{\bibfnamefont{X.}~\bibnamefont{Zotos}},
  \bibinfo{author}{\bibfnamefont{F.}~\bibnamefont{Naef}}, \bibnamefont{and}
  \bibinfo{author}{\bibfnamefont{P.}~\bibnamefont{Prelovsek}},
  \bibinfo{journal}{Phys. Rev. B} \textbf{\bibinfo{volume}{55}},
  \bibinfo{pages}{11029} (\bibinfo{year}{1997}).

\bibitem[{\citenamefont{Heidrich-Meisner
  et~al.}(2003)\citenamefont{Heidrich-Meisner, Honecker, Cabra, and
  Brenig}}]{Heidrich2003}
\bibinfo{author}{\bibfnamefont{F.}~\bibnamefont{Heidrich-Meisner}},
  \bibinfo{author}{\bibfnamefont{A.}~\bibnamefont{Honecker}},
  \bibinfo{author}{\bibfnamefont{D.}~\bibnamefont{Cabra}}, \bibnamefont{and}
  \bibinfo{author}{\bibfnamefont{W.}~\bibnamefont{Brenig}},
  \bibinfo{journal}{Phys. Rev. B} \textbf{\bibinfo{volume}{68}},
  \bibinfo{pages}{134436} (\bibinfo{year}{2003}).

\bibitem[{\citenamefont{Kl\"umper and Sakai}(2002)}]{Kluemper2002}
\bibinfo{author}{\bibfnamefont{A.}~\bibnamefont{Kl\"umper}} \bibnamefont{and}
  \bibinfo{author}{\bibfnamefont{K.}~\bibnamefont{Sakai}}, \bibinfo{journal}{J.
  Phys. A} \textbf{\bibinfo{volume}{35}}, \bibinfo{pages}{2173}
  (\bibinfo{year}{2002}).

\bibitem[{\citenamefont{Buchanan}(2005)}]{Buchanan2005}
\bibinfo{author}{\bibfnamefont{M.}~\bibnamefont{Buchanan}},
  \bibinfo{journal}{Nature Phys.} \textbf{\bibinfo{volume}{1}},
  \bibinfo{pages}{71} (\bibinfo{year}{2005}).

\bibitem[{\citenamefont{Casati et~al.}(1984)\citenamefont{Casati, Ford,
  Vivaldi, and Visscher}}]{Casati1984}
\bibinfo{author}{\bibfnamefont{G.}~\bibnamefont{Casati}},
  \bibinfo{author}{\bibfnamefont{J.}~\bibnamefont{Ford}},
  \bibinfo{author}{\bibfnamefont{F.}~\bibnamefont{Vivaldi}}, \bibnamefont{and}
  \bibinfo{author}{\bibfnamefont{W.}~\bibnamefont{Visscher}},
  \bibinfo{journal}{Phys. Rev. Lett.} \textbf{\bibinfo{volume}{52}},
  \bibinfo{pages}{1861} (\bibinfo{year}{1984}).

\bibitem[{\citenamefont{Prosen}(1998)}]{Prosen1998}
\bibinfo{author}{\bibfnamefont{T.}~\bibnamefont{Prosen}},
  \bibinfo{journal}{Phys. Rev. Lett.} \textbf{\bibinfo{volume}{80}},
  \bibinfo{pages}{1808} (\bibinfo{year}{1998}).

\bibitem[{\citenamefont{Vollmer}(2002)}]{Vollmer2002}
\bibinfo{author}{\bibfnamefont{J.}~\bibnamefont{Vollmer}},
  \bibinfo{journal}{Phys. Rep.} \textbf{\bibinfo{volume}{372}},
  \bibinfo{pages}{131} (\bibinfo{year}{2002}).

\bibitem[{\citenamefont{Larralde et~al.}(2003)\citenamefont{Larralde, Leyvraz,
  and Mejia-Monasterio}}]{Larralde2003}
\bibinfo{author}{\bibfnamefont{H.}~\bibnamefont{Larralde}},
  \bibinfo{author}{\bibfnamefont{F.}~\bibnamefont{Leyvraz}}, \bibnamefont{and}
  \bibinfo{author}{\bibfnamefont{C.}~\bibnamefont{Mejia-Monasterio}},
  \bibinfo{journal}{J. Stat. Phys.} \textbf{\bibinfo{volume}{113}},
  \bibinfo{pages}{197} (\bibinfo{year}{2003}).

\bibitem[{\citenamefont{Li et~al.}(2004)\citenamefont{Li, Casati, Wang, and
  Prosen}}]{Li2004}
\bibinfo{author}{\bibfnamefont{B.}~\bibnamefont{Li}},
  \bibinfo{author}{\bibfnamefont{G.}~\bibnamefont{Casati}},
  \bibinfo{author}{\bibfnamefont{J.}~\bibnamefont{Wang}}, \bibnamefont{and}
  \bibinfo{author}{\bibfnamefont{T.}~\bibnamefont{Prosen}},
  \bibinfo{journal}{Phys. Rev. Lett.} \textbf{\bibinfo{volume}{92}},
  \bibinfo{pages}{254301} (\bibinfo{year}{2004}).

\bibitem[{\citenamefont{{Prosen} and {Robnik}}(1992)}]{Prosen1992}
\bibinfo{author}{\bibfnamefont{T.}~\bibnamefont{{Prosen}}} \bibnamefont{and}
  \bibinfo{author}{\bibfnamefont{M.}~\bibnamefont{{Robnik}}},
  \bibinfo{journal}{J. Phys. A} \textbf{\bibinfo{volume}{25}},
  \bibinfo{pages}{3449} (\bibinfo{year}{1992}).

\bibitem[{\citenamefont{Hu et~al.}(1998)\citenamefont{Hu, Li, and
  Zhao}}]{Hu1998}
\bibinfo{author}{\bibfnamefont{B.}~\bibnamefont{Hu}},
  \bibinfo{author}{\bibfnamefont{B.}~\bibnamefont{Li}}, \bibnamefont{and}
  \bibinfo{author}{\bibfnamefont{H.}~\bibnamefont{Zhao}},
  \bibinfo{journal}{Phys. Rev. E} \textbf{\bibinfo{volume}{57}},
  \bibinfo{pages}{2992} (\bibinfo{year}{1998}).

\bibitem[{\citenamefont{Garrido et~al.}(2001)\citenamefont{Garrido, Hurtado,
  and Nadrowski}}]{Garrido2001}
\bibinfo{author}{\bibfnamefont{P.}~\bibnamefont{Garrido}},
  \bibinfo{author}{\bibfnamefont{P.}~\bibnamefont{Hurtado}}, \bibnamefont{and}
  \bibinfo{author}{\bibfnamefont{B.}~\bibnamefont{Nadrowski}},
  \bibinfo{journal}{Phys.\ Rev.\ Lett.} \textbf{\bibinfo{volume}{86}},
  \bibinfo{pages}{5486} (\bibinfo{year}{2001}).

\bibitem[{\citenamefont{Lepri et~al.}(1997)\citenamefont{Lepri, Livi, and
  Politi}}]{Lepri1997}
\bibinfo{author}{\bibfnamefont{S.}~\bibnamefont{Lepri}},
  \bibinfo{author}{\bibfnamefont{R.}~\bibnamefont{Livi}}, \bibnamefont{and}
  \bibinfo{author}{\bibfnamefont{A.}~\bibnamefont{Politi}},
  \bibinfo{journal}{Phys. Rev. Lett.} \textbf{\bibinfo{volume}{78}},
  \bibinfo{pages}{1896} (\bibinfo{year}{1997}).

\bibitem[{\citenamefont{Lepri et~al.}(1998)\citenamefont{Lepri, Livi, and
  Politi}}]{Lepri1998}
\bibinfo{author}{\bibfnamefont{S.}~\bibnamefont{Lepri}},
  \bibinfo{author}{\bibfnamefont{R.}~\bibnamefont{Livi}}, \bibnamefont{and}
  \bibinfo{author}{\bibfnamefont{A.}~\bibnamefont{Politi}},
  \bibinfo{journal}{Europhys. Lett.} \textbf{\bibinfo{volume}{43}},
  \bibinfo{pages}{271} (\bibinfo{year}{1998}).

\bibitem[{\citenamefont{Savin et~al.}(2002)\citenamefont{Savin, Tsironis, and
  Zolotaryuk}}]{Savin2002}
\bibinfo{author}{\bibfnamefont{A.}~\bibnamefont{Savin}},
  \bibinfo{author}{\bibfnamefont{G.}~\bibnamefont{Tsironis}}, \bibnamefont{and}
  \bibinfo{author}{\bibfnamefont{A.}~\bibnamefont{Zolotaryuk}},
  \bibinfo{journal}{Phys.\ Rev.\ Lett.} \textbf{\bibinfo{volume}{88}},
  \bibinfo{pages}{154301} (\bibinfo{year}{2002}).

\bibitem[{\citenamefont{Dhar}(2001)}]{Dhar2001}
\bibinfo{author}{\bibfnamefont{A.}~\bibnamefont{Dhar}}, \bibinfo{journal}{Phys.
  Rev. Lett.} \textbf{\bibinfo{volume}{86}}, \bibinfo{pages}{5882}
  (\bibinfo{year}{2001}).

\bibitem[{\citenamefont{Chiu et~al.}(2005)\citenamefont{Chiu, Deshpande,
  Postma, Lau, Miko, Forro, and Bockrath}}]{Chiu2005}
\bibinfo{author}{\bibfnamefont{H.-Y.} \bibnamefont{Chiu}},
  \bibinfo{author}{\bibfnamefont{V.~V.} \bibnamefont{Deshpande}},
  \bibinfo{author}{\bibfnamefont{H.~W.~C.} \bibnamefont{Postma}},
  \bibinfo{author}{\bibfnamefont{C.~N.} \bibnamefont{Lau}},
  \bibinfo{author}{\bibfnamefont{C.}~\bibnamefont{Miko}},
  \bibinfo{author}{\bibfnamefont{L.}~\bibnamefont{Forro}}, \bibnamefont{and}
  \bibinfo{author}{\bibfnamefont{M.}~\bibnamefont{Bockrath}},
  \bibinfo{journal}{Phys. Rev. Lett.} \textbf{\bibinfo{volume}{95}},
  \bibinfo{pages}{226101} (\bibinfo{year}{2005}).

\bibitem[{\citenamefont{Yu et~al.}(2005)\citenamefont{Yu, Shi, Yao, Li, and
  Majumdar}}]{Yu2005}
\bibinfo{author}{\bibfnamefont{C.}~\bibnamefont{Yu}},
  \bibinfo{author}{\bibfnamefont{L.}~\bibnamefont{Shi}},
  \bibinfo{author}{\bibfnamefont{Z.}~\bibnamefont{Yao}},
  \bibinfo{author}{\bibfnamefont{D.}~\bibnamefont{Li}}, \bibnamefont{and}
  \bibinfo{author}{\bibfnamefont{A.}~\bibnamefont{Majumdar}},
  \bibinfo{journal}{Nano Lett.} \textbf{\bibinfo{volume}{5}},
  \bibinfo{pages}{1842} (\bibinfo{year}{2005}).

\bibitem[{\citenamefont{Fujii et~al.}(2005)\citenamefont{Fujii, Zhang, Xie,
  Ago, Takahashi, Ikuta, Abe, and Shimizu}}]{Fujii2005}
\bibinfo{author}{\bibfnamefont{M.}~\bibnamefont{Fujii}},
  \bibinfo{author}{\bibfnamefont{X.}~\bibnamefont{Zhang}},
  \bibinfo{author}{\bibfnamefont{H.}~\bibnamefont{Xie}},
  \bibinfo{author}{\bibfnamefont{H.}~\bibnamefont{Ago}},
  \bibinfo{author}{\bibfnamefont{K.}~\bibnamefont{Takahashi}},
  \bibinfo{author}{\bibfnamefont{T.}~\bibnamefont{Ikuta}},
  \bibinfo{author}{\bibfnamefont{H.}~\bibnamefont{Abe}}, \bibnamefont{and}
  \bibinfo{author}{\bibfnamefont{T.}~\bibnamefont{Shimizu}},
  \bibinfo{journal}{Phys. Rev. Lett.} \textbf{\bibinfo{volume}{95}},
  \bibinfo{pages}{065502} (\bibinfo{year}{2005}).

\bibitem[{\citenamefont{Sologubenko
  et~al.}(2000{\natexlab{a}})\citenamefont{Sologubenko, Giann\'o, Ott,
  Ammerahl, and Revcolevschi}}]{Sologubenko2000I}
\bibinfo{author}{\bibfnamefont{A.}~\bibnamefont{Sologubenko}},
  \bibinfo{author}{\bibfnamefont{K.}~\bibnamefont{Giann\'o}},
  \bibinfo{author}{\bibfnamefont{H.}~\bibnamefont{Ott}},
  \bibinfo{author}{\bibfnamefont{U.}~\bibnamefont{Ammerahl}}, \bibnamefont{and}
  \bibinfo{author}{\bibfnamefont{A.}~\bibnamefont{Revcolevschi}},
  \bibinfo{journal}{Phy. Rev. Lett.} \textbf{\bibinfo{volume}{84}},
  \bibinfo{pages}{2714} (\bibinfo{year}{2000}{\natexlab{a}}).

\bibitem[{\citenamefont{Sologubenko
  et~al.}(2000{\natexlab{b}})\citenamefont{Sologubenko, Felder, Giann\`o, Ott,
  Vietkine, and Revcolevschi}}]{Sologubenko2000II}
\bibinfo{author}{\bibfnamefont{A.}~\bibnamefont{Sologubenko}},
  \bibinfo{author}{\bibfnamefont{E.}~\bibnamefont{Felder}},
  \bibinfo{author}{\bibfnamefont{K.}~\bibnamefont{Giann\`o}},
  \bibinfo{author}{\bibfnamefont{H.}~\bibnamefont{Ott}},
  \bibinfo{author}{\bibfnamefont{A.}~\bibnamefont{Vietkine}}, \bibnamefont{and}
  \bibinfo{author}{\bibfnamefont{A.}~\bibnamefont{Revcolevschi}},
  \bibinfo{journal}{Phys. Rev. B} \textbf{\bibinfo{volume}{62}},
  \bibinfo{pages}{R6108} (\bibinfo{year}{2000}{\natexlab{b}}).

\bibitem[{\citenamefont{Sologubenko
  et~al.}(2000{\natexlab{c}})\citenamefont{Sologubenko, Giann\'o, Ott,
  Ammerahl, Revcolevschi, Brewer, and Thomson}}]{Sologubenko2000III}
\bibinfo{author}{\bibfnamefont{A.}~\bibnamefont{Sologubenko}},
  \bibinfo{author}{\bibfnamefont{K.}~\bibnamefont{Giann\'o}},
  \bibinfo{author}{\bibfnamefont{H.}~\bibnamefont{Ott}},
  \bibinfo{author}{\bibfnamefont{U.}~\bibnamefont{Ammerahl}},
  \bibinfo{author}{\bibfnamefont{A.}~\bibnamefont{Revcolevschi}},
  \bibinfo{author}{\bibfnamefont{D.}~\bibnamefont{Brewer}}, \bibnamefont{and}
  \bibinfo{author}{\bibfnamefont{A.}~\bibnamefont{Thomson}},
  \bibinfo{journal}{Physica B} \textbf{\bibinfo{volume}{284}},
  \bibinfo{pages}{1595} (\bibinfo{year}{2000}{\natexlab{c}}).

\bibitem[{\citenamefont{Sologubenko et~al.}(2003)\citenamefont{Sologubenko,
  Kazakov, Ott, Asano, and Ajiro}}]{Sologubenko2003}
\bibinfo{author}{\bibfnamefont{A.}~\bibnamefont{Sologubenko}},
  \bibinfo{author}{\bibfnamefont{S.}~\bibnamefont{Kazakov}},
  \bibinfo{author}{\bibfnamefont{H.}~\bibnamefont{Ott}},
  \bibinfo{author}{\bibfnamefont{T.}~\bibnamefont{Asano}}, \bibnamefont{and}
  \bibinfo{author}{\bibfnamefont{Y.}~\bibnamefont{Ajiro}},
  \bibinfo{journal}{Phys. Rev. B} \textbf{\bibinfo{volume}{68}},
  \bibinfo{pages}{094432} (\bibinfo{year}{2003}).

\bibitem[{\citenamefont{Hess et~al.}(2001)\citenamefont{Hess, Baumann,
  Ammerahl, B\"uchner, Heidrich-Meisner, Brenig, and Revcolevschi}}]{Hess2001}
\bibinfo{author}{\bibfnamefont{C.}~\bibnamefont{Hess}},
  \bibinfo{author}{\bibfnamefont{C.}~\bibnamefont{Baumann}},
  \bibinfo{author}{\bibfnamefont{U.}~\bibnamefont{Ammerahl}},
  \bibinfo{author}{\bibfnamefont{B.}~\bibnamefont{B\"uchner}},
  \bibinfo{author}{\bibfnamefont{F.}~\bibnamefont{Heidrich-Meisner}},
  \bibinfo{author}{\bibfnamefont{W.}~\bibnamefont{Brenig}}, \bibnamefont{and}
  \bibinfo{author}{\bibfnamefont{A.}~\bibnamefont{Revcolevschi}},
  \bibinfo{journal}{Phys. Rev. B} \textbf{\bibinfo{volume}{64}},
  \bibinfo{pages}{184305} (\bibinfo{year}{2001}).

\bibitem[{\citenamefont{Hess et~al.}(2004)\citenamefont{Hess, ElHaes,
  B\"uchner, Ammerahl, Hucker, and Revcolevschi}}]{Hess2004}
\bibinfo{author}{\bibfnamefont{C.}~\bibnamefont{Hess}},
  \bibinfo{author}{\bibfnamefont{H.}~\bibnamefont{ElHaes}},
  \bibinfo{author}{\bibfnamefont{B.}~\bibnamefont{B\"uchner}},
  \bibinfo{author}{\bibfnamefont{U.}~\bibnamefont{Ammerahl}},
  \bibinfo{author}{\bibfnamefont{M.}~\bibnamefont{Hucker}}, \bibnamefont{and}
  \bibinfo{author}{\bibfnamefont{A.}~\bibnamefont{Revcolevschi}},
  \bibinfo{journal}{Phys. Rev. Lett.} \textbf{\bibinfo{volume}{93}},
  \bibinfo{pages}{027005} (\bibinfo{year}{2004}).

\bibitem[{\citenamefont{Ribeiro et~al.}(2005)\citenamefont{Ribeiro, Hess,
  Reutler, Roth, and B\"uchner}}]{Ribeiro2005}
\bibinfo{author}{\bibfnamefont{P.}~\bibnamefont{Ribeiro}},
  \bibinfo{author}{\bibfnamefont{C.}~\bibnamefont{Hess}},
  \bibinfo{author}{\bibfnamefont{P.}~\bibnamefont{Reutler}},
  \bibinfo{author}{\bibfnamefont{G.}~\bibnamefont{Roth}}, \bibnamefont{and}
  \bibinfo{author}{\bibfnamefont{B.}~\bibnamefont{B\"uchner}},
  \bibinfo{journal}{J. Magn. Magn. Mater.} \textbf{\bibinfo{volume}{290-291}},
  \bibinfo{pages}{334} (\bibinfo{year}{2005}).

\bibitem[{\citenamefont{Zotos and Prelovsek}(1996)}]{Zotos1996}
\bibinfo{author}{\bibfnamefont{X.}~\bibnamefont{Zotos}} \bibnamefont{and}
  \bibinfo{author}{\bibfnamefont{P.}~\bibnamefont{Prelovsek}},
  \bibinfo{journal}{Phys. Rev. B} \textbf{\bibinfo{volume}{53}},
  \bibinfo{pages}{983} (\bibinfo{year}{1996}).

\bibitem[{\citenamefont{Heidrich-Meisner
  et~al.}(2002)\citenamefont{Heidrich-Meisner, Honecker, Cabra, and
  Brenig}}]{Heidrich2002}
\bibinfo{author}{\bibfnamefont{F.}~\bibnamefont{Heidrich-Meisner}},
  \bibinfo{author}{\bibfnamefont{A.}~\bibnamefont{Honecker}},
  \bibinfo{author}{\bibfnamefont{D.}~\bibnamefont{Cabra}}, \bibnamefont{and}
  \bibinfo{author}{\bibfnamefont{W.}~\bibnamefont{Brenig}},
  \bibinfo{journal}{Phys. Rev. B} \textbf{\bibinfo{volume}{66}},
  \bibinfo{pages}{140406} (\bibinfo{year}{2002}).

\bibitem[{\citenamefont{Heidrich-Meissner
  et~al.}(2004)\citenamefont{Heidrich-Meissner, Honecker, Cabra, and
  Brenig}}]{Heidrich2004II}
\bibinfo{author}{\bibfnamefont{F.}~\bibnamefont{Heidrich-Meissner}},
  \bibinfo{author}{\bibfnamefont{A.}~\bibnamefont{Honecker}},
  \bibinfo{author}{\bibfnamefont{D.}~\bibnamefont{Cabra}}, \bibnamefont{and}
  \bibinfo{author}{\bibfnamefont{W.}~\bibnamefont{Brenig}},
  \bibinfo{journal}{J. Mag. Mag. Mater.} \textbf{\bibinfo{volume}{272-276}},
  \bibinfo{pages}{890} (\bibinfo{year}{2004}).

\bibitem[{\citenamefont{Heidrich-Meisner
  et~al.}(2005)\citenamefont{Heidrich-Meisner, Honecker, Cabra, and
  Brenig}}]{Heidrich2005II}
\bibinfo{author}{\bibfnamefont{F.}~\bibnamefont{Heidrich-Meisner}},
  \bibinfo{author}{\bibfnamefont{A.}~\bibnamefont{Honecker}},
  \bibinfo{author}{\bibfnamefont{D.}~\bibnamefont{Cabra}}, \bibnamefont{and}
  \bibinfo{author}{\bibfnamefont{W.}~\bibnamefont{Brenig}},
  \bibinfo{journal}{Physica B} \textbf{\bibinfo{volume}{359-361}},
  \bibinfo{pages}{1394} (\bibinfo{year}{2005}).

\bibitem[{\citenamefont{Kl\"umper and Johnston}(2000)}]{Kluemper2000}
\bibinfo{author}{\bibfnamefont{A.}~\bibnamefont{Kl\"umper}} \bibnamefont{and}
  \bibinfo{author}{\bibfnamefont{D.}~\bibnamefont{Johnston}},
  \bibinfo{journal}{Phys. Rev. Lett.} \textbf{\bibinfo{volume}{84}},
  \bibinfo{pages}{4701} (\bibinfo{year}{2000}).

\bibitem[{\citenamefont{Jung et~al.}(2006)\citenamefont{Jung, Helmes, and
  Rosch}}]{Jung2006}
\bibinfo{author}{\bibfnamefont{P.}~\bibnamefont{Jung}},
  \bibinfo{author}{\bibfnamefont{R.}~\bibnamefont{Helmes}}, \bibnamefont{and}
  \bibinfo{author}{\bibfnamefont{A.}~\bibnamefont{Rosch}},
  \bibinfo{journal}{Phys. Rev. Lett.} \textbf{\bibinfo{volume}{96}},
  \bibinfo{pages}{067202} (\bibinfo{year}{2006}).

\bibitem[{\citenamefont{Gemmer et~al.}(2006)\citenamefont{Gemmer, Steinigeweg,
  and Michel}}]{Gemmer2006II}
\bibinfo{author}{\bibfnamefont{J.}~\bibnamefont{Gemmer}},
  \bibinfo{author}{\bibfnamefont{R.}~\bibnamefont{Steinigeweg}},
  \bibnamefont{and} \bibinfo{author}{\bibfnamefont{M.}~\bibnamefont{Michel}},
  \bibinfo{journal}{Phys. Rev. B} \textbf{\bibinfo{volume}{73}},
  \bibinfo{pages}{104302} (\bibinfo{year}{2006}).

\bibitem[{\citenamefont{Michel et~al.}(2005{\natexlab{a}})\citenamefont{Michel,
  Gemmer, and Mahler}}]{Michel2005}
\bibinfo{author}{\bibfnamefont{M.}~\bibnamefont{Michel}},
  \bibinfo{author}{\bibfnamefont{J.}~\bibnamefont{Gemmer}}, \bibnamefont{and}
  \bibinfo{author}{\bibfnamefont{G.}~\bibnamefont{Mahler}},
  \bibinfo{journal}{Physica E} \textbf{\bibinfo{volume}{29}},
  \bibinfo{pages}{129} (\bibinfo{year}{2005}{\natexlab{a}}).

\bibitem[{\citenamefont{Saito et~al.}(1996{\natexlab{a}})\citenamefont{Saito,
  Takesue, and Miyashita}}]{Saito1996}
\bibinfo{author}{\bibfnamefont{K.}~\bibnamefont{Saito}},
  \bibinfo{author}{\bibfnamefont{S.}~\bibnamefont{Takesue}}, \bibnamefont{and}
  \bibinfo{author}{\bibfnamefont{S.}~\bibnamefont{Miyashita}},
  \bibinfo{journal}{Phys. Rev. E} \textbf{\bibinfo{volume}{54}},
  \bibinfo{pages}{2404} (\bibinfo{year}{1996}{\natexlab{a}}).

\bibitem[{\citenamefont{Saito et~al.}(1996{\natexlab{b}})\citenamefont{Saito,
  Takesue, and Miyashita}}]{Saito1996II}
\bibinfo{author}{\bibfnamefont{K.}~\bibnamefont{Saito}},
  \bibinfo{author}{\bibfnamefont{S.}~\bibnamefont{Takesue}}, \bibnamefont{and}
  \bibinfo{author}{\bibfnamefont{S.}~\bibnamefont{Miyashita}},
  \bibinfo{journal}{J. Phys. Soc. Jpn.} \textbf{\bibinfo{volume}{65}},
  \bibinfo{pages}{1243} (\bibinfo{year}{1996}{\natexlab{b}}).

\bibitem[{\citenamefont{Saito et~al.}(2000)\citenamefont{Saito, Takesue, and
  Miyashita}}]{Saito2000}
\bibinfo{author}{\bibfnamefont{K.}~\bibnamefont{Saito}},
  \bibinfo{author}{\bibfnamefont{S.}~\bibnamefont{Takesue}}, \bibnamefont{and}
  \bibinfo{author}{\bibfnamefont{S.}~\bibnamefont{Miyashita}},
  \bibinfo{journal}{Phys. Rev. E} \textbf{\bibinfo{volume}{61}},
  \bibinfo{pages}{2397} (\bibinfo{year}{2000}).

\bibitem[{\citenamefont{Saito and Miyashita}(2002)}]{Saito2002}
\bibinfo{author}{\bibfnamefont{K.}~\bibnamefont{Saito}} \bibnamefont{and}
  \bibinfo{author}{\bibfnamefont{S.}~\bibnamefont{Miyashita}},
  \bibinfo{journal}{J. Phys. Soc. Jpn.} \textbf{\bibinfo{volume}{71}},
  \bibinfo{pages}{2485} (\bibinfo{year}{2002}).

\bibitem[{\citenamefont{Saito}(2003)}]{Saito2003}
\bibinfo{author}{\bibfnamefont{K.}~\bibnamefont{Saito}},
  \bibinfo{journal}{Europhys. Lett.} \textbf{\bibinfo{volume}{61}},
  \bibinfo{pages}{34} (\bibinfo{year}{2003}).

\bibitem[{\citenamefont{Michel et~al.}(2003)\citenamefont{Michel, Hartmann,
  Gemmer, and Mahler}}]{Michel2003}
\bibinfo{author}{\bibfnamefont{M.}~\bibnamefont{Michel}},
  \bibinfo{author}{\bibfnamefont{M.}~\bibnamefont{Hartmann}},
  \bibinfo{author}{\bibfnamefont{J.}~\bibnamefont{Gemmer}}, \bibnamefont{and}
  \bibinfo{author}{\bibfnamefont{G.}~\bibnamefont{Mahler}},
  \bibinfo{journal}{Eur. Phys. J. B} \textbf{\bibinfo{volume}{34}},
  \bibinfo{pages}{325} (\bibinfo{year}{2003}).

\bibitem[{\citenamefont{Lecomte et~al.}(2005)\citenamefont{Lecomte, R\'acz, and
  van Wijland}}]{Lecomte2005}
\bibinfo{author}{\bibfnamefont{V.}~\bibnamefont{Lecomte}},
  \bibinfo{author}{\bibfnamefont{Z.}~\bibnamefont{R\'acz}}, \bibnamefont{and}
  \bibinfo{author}{\bibfnamefont{F.}~\bibnamefont{van Wijland}},
  \bibinfo{journal}{J. Stat. Mech.} \textbf{\bibinfo{volume}{2}}
  (\bibinfo{year}{2005}).

\bibitem[{\citenamefont{Landauer}(1957)}]{Landauer1957}
\bibinfo{author}{\bibfnamefont{R.}~\bibnamefont{Landauer}},
  \bibinfo{journal}{IBM J. Res. Dev.} \textbf{\bibinfo{volume}{1}},
  \bibinfo{pages}{223} (\bibinfo{year}{1957}).

\bibitem[{\citenamefont{Segal et~al.}(2003)\citenamefont{Segal, Nitzana, and
  H\"anggi}}]{Segal2003}
\bibinfo{author}{\bibfnamefont{D.}~\bibnamefont{Segal}},
  \bibinfo{author}{\bibfnamefont{A.}~\bibnamefont{Nitzana}}, \bibnamefont{and}
  \bibinfo{author}{\bibfnamefont{P.}~\bibnamefont{H\"anggi}},
  \bibinfo{journal}{J. Chem. Phys.} \textbf{\bibinfo{volume}{119}},
  \bibinfo{pages}{6840} (\bibinfo{year}{2003}).

\bibitem[{\citenamefont{Segal and Nitzan}(2005{\natexlab{a}})}]{Segal2005I}
\bibinfo{author}{\bibfnamefont{D.}~\bibnamefont{Segal}} \bibnamefont{and}
  \bibinfo{author}{\bibfnamefont{A.}~\bibnamefont{Nitzan}},
  \bibinfo{journal}{Phys. Rev. Lett.} \textbf{\bibinfo{volume}{94}},
  \bibinfo{pages}{034301} (\bibinfo{year}{2005}{\natexlab{a}}).

\bibitem[{\citenamefont{Segal and Nitzan}(2005{\natexlab{b}})}]{Segal2005II}
\bibinfo{author}{\bibfnamefont{D.}~\bibnamefont{Segal}} \bibnamefont{and}
  \bibinfo{author}{\bibfnamefont{A.}~\bibnamefont{Nitzan}},
  \bibinfo{journal}{J. Chem. Phys.} \textbf{\bibinfo{volume}{122}},
  \bibinfo{pages}{194704} (\bibinfo{year}{2005}{\natexlab{b}}).

\bibitem[{\citenamefont{Antal et~al.}(1997)\citenamefont{Antal, Racz, and
  Sasvari}}]{Antal1997}
\bibinfo{author}{\bibfnamefont{T.}~\bibnamefont{Antal}},
  \bibinfo{author}{\bibfnamefont{Z.}~\bibnamefont{Racz}}, \bibnamefont{and}
  \bibinfo{author}{\bibfnamefont{L.}~\bibnamefont{Sasvari}},
  \bibinfo{journal}{Phys. Rev. Lett.} \textbf{\bibinfo{volume}{78}},
  \bibinfo{pages}{167} (\bibinfo{year}{1997}).

\bibitem[{\citenamefont{Antal et~al.}(1999)\citenamefont{Antal, Racz, Rakos,
  and Sch\"utz}}]{Antal1999}
\bibinfo{author}{\bibfnamefont{T.}~\bibnamefont{Antal}},
  \bibinfo{author}{\bibfnamefont{Z.}~\bibnamefont{Racz}},
  \bibinfo{author}{\bibfnamefont{A.}~\bibnamefont{Rakos}}, \bibnamefont{and}
  \bibinfo{author}{\bibfnamefont{G.}~\bibnamefont{Sch\"utz}},
  \bibinfo{journal}{Phys. Rev. E} \textbf{\bibinfo{volume}{59}},
  \bibinfo{pages}{4912} (\bibinfo{year}{1999}).

\bibitem[{\citenamefont{Eisler et~al.}(2003)\citenamefont{Eisler, Racz, and van
  Wijland}}]{Eisler2003}
\bibinfo{author}{\bibfnamefont{V.}~\bibnamefont{Eisler}},
  \bibinfo{author}{\bibfnamefont{Z.}~\bibnamefont{Racz}}, \bibnamefont{and}
  \bibinfo{author}{\bibfnamefont{F.}~\bibnamefont{van Wijland}},
  \bibinfo{journal}{Phys. Rev. E} \textbf{\bibinfo{volume}{67}},
  \bibinfo{pages}{056129} (\bibinfo{year}{2003}).

\bibitem[{\citenamefont{Gobert et~al.}(2005)\citenamefont{Gobert, Kollath,
  Schollw\"ock, and Sch\"utz}}]{Gobert2005}
\bibinfo{author}{\bibfnamefont{D.}~\bibnamefont{Gobert}},
  \bibinfo{author}{\bibfnamefont{C.}~\bibnamefont{Kollath}},
  \bibinfo{author}{\bibfnamefont{U.}~\bibnamefont{Schollw\"ock}},
  \bibnamefont{and} \bibinfo{author}{\bibfnamefont{G.}~\bibnamefont{Sch\"utz}},
  \bibinfo{journal}{Phys. Rev. E} \textbf{\bibinfo{volume}{71}},
  \bibinfo{pages}{036102} (\bibinfo{year}{2005}).

\bibitem[{\citenamefont{Mejia-Monasterio
  et~al.}(2005)\citenamefont{Mejia-Monasterio, Prosen, and
  Casati}}]{MejiaMonasterio2005}
\bibinfo{author}{\bibfnamefont{C.}~\bibnamefont{Mejia-Monasterio}},
  \bibinfo{author}{\bibfnamefont{T.}~\bibnamefont{Prosen}}, \bibnamefont{and}
  \bibinfo{author}{\bibfnamefont{G.}~\bibnamefont{Casati}},
  \bibinfo{journal}{Europhys. Lett.} \textbf{\bibinfo{volume}{72}},
  \bibinfo{pages}{520} (\bibinfo{year}{2005}).

\bibitem[{\citenamefont{Haake}(2004)}]{Haake2004}
\bibinfo{author}{\bibfnamefont{F.}~\bibnamefont{Haake}},
  \emph{\bibinfo{title}{Quantum Signatures of Chaos}}
  (\bibinfo{publisher}{Springer}, \bibinfo{address}{Berlin, Heidelberg,
  New-York}, \bibinfo{year}{2004}), \bibinfo{edition}{2nd} ed.

\bibitem[{\citenamefont{Steinigeweg et~al.}(2006)\citenamefont{Steinigeweg,
  Gemmer, and Michel}}]{Steinigeweg2006I}
\bibinfo{author}{\bibfnamefont{R.}~\bibnamefont{Steinigeweg}},
  \bibinfo{author}{\bibfnamefont{J.}~\bibnamefont{Gemmer}}, \bibnamefont{and}
  \bibinfo{author}{\bibfnamefont{M.}~\bibnamefont{Michel}},
  \bibinfo{journal}{Europhys. Lett.} \textbf{\bibinfo{volume}{75}},
  \bibinfo{pages}{406} (\bibinfo{year}{2006}).

\bibitem[{\citenamefont{Choquard}(1963)}]{Choquard1963}
\bibinfo{author}{\bibfnamefont{P.}~\bibnamefont{Choquard}},
  \bibinfo{journal}{Helv. Phys. Acta} \textbf{\bibinfo{volume}{36}},
  \bibinfo{pages}{415} (\bibinfo{year}{1963}).

\bibitem[{\citenamefont{Gemmer et~al.}(2004)\citenamefont{Gemmer, Michel, and
  Mahler}}]{Gemmer2004}
\bibinfo{author}{\bibfnamefont{J.}~\bibnamefont{Gemmer}},
  \bibinfo{author}{\bibfnamefont{M.}~\bibnamefont{Michel}}, \bibnamefont{and}
  \bibinfo{author}{\bibfnamefont{G.}~\bibnamefont{Mahler}},
  \emph{\bibinfo{title}{Quantum Thermodynamics: Emergence of Thermodynamic
  Behavior within Composite Quantum Systems}}, LNP657
  (\bibinfo{publisher}{Springer}, \bibinfo{address}{Berlin, Heidelberg,
  New-York}, \bibinfo{year}{2004}).

\bibitem[{\citenamefont{Schack and Caves}(2000)}]{Schack2000}
\bibinfo{author}{\bibfnamefont{R.}~\bibnamefont{Schack}} \bibnamefont{and}
  \bibinfo{author}{\bibfnamefont{M.}~\bibnamefont{Caves}}, \bibinfo{journal}{J.
  Mod. Opt.} \textbf{\bibinfo{volume}{47}}, \bibinfo{pages}{387}
  (\bibinfo{year}{2000}).

\bibitem[{\citenamefont{Lindblad}(1976)}]{Lindblad1976}
\bibinfo{author}{\bibfnamefont{G.}~\bibnamefont{Lindblad}},
  \bibinfo{journal}{Commun. Math. Phys.} \textbf{\bibinfo{volume}{48}},
  \bibinfo{pages}{119} (\bibinfo{year}{1976}).

\bibitem[{\citenamefont{Breuer and Petruccione}(2002)}]{Breuer2002}
\bibinfo{author}{\bibfnamefont{H.-P.} \bibnamefont{Breuer}} \bibnamefont{and}
  \bibinfo{author}{\bibfnamefont{F.}~\bibnamefont{Petruccione}},
  \emph{\bibinfo{title}{The Theory of Open Quantum Systems}}
  (\bibinfo{publisher}{Oxford University Press}, \bibinfo{address}{Oxford},
  \bibinfo{year}{2002}).

\bibitem[{\citenamefont{Tarasov}(2002{\natexlab{a}})}]{Tarasov2002}
\bibinfo{author}{\bibfnamefont{V.}~\bibnamefont{Tarasov}},
  \bibinfo{journal}{Phys. Rev. E} \textbf{\bibinfo{volume}{66}},
  \bibinfo{pages}{056116} (\bibinfo{year}{2002}{\natexlab{a}}).

\bibitem[{\citenamefont{Tarasov}(2002{\natexlab{b}})}]{Tarasov2002II}
\bibinfo{author}{\bibfnamefont{V.}~\bibnamefont{Tarasov}},
  \bibinfo{journal}{Phys. Lett. A} \textbf{\bibinfo{volume}{299}},
  \bibinfo{pages}{173} (\bibinfo{year}{2002}{\natexlab{b}}).

\bibitem[{\citenamefont{Mukamel}(2003)}]{Mukamel2003}
\bibinfo{author}{\bibfnamefont{S.}~\bibnamefont{Mukamel}},
  \bibinfo{journal}{Phys. Rev. E} \textbf{\bibinfo{volume}{68}},
  \bibinfo{pages}{021111} (\bibinfo{year}{2003}).

\bibitem[{\citenamefont{Caves}(1999)}]{Caves1999}
\bibinfo{author}{\bibfnamefont{C.}~\bibnamefont{Caves}}, \bibinfo{journal}{J.
  Supercond.} \textbf{\bibinfo{volume}{12}}, \bibinfo{pages}{707}
  (\bibinfo{year}{1999}).

\bibitem[{\citenamefont{Michel et~al.}(2004)\citenamefont{Michel, Gemmer, and
  Mahler}}]{Michel2004}
\bibinfo{author}{\bibfnamefont{M.}~\bibnamefont{Michel}},
  \bibinfo{author}{\bibfnamefont{J.}~\bibnamefont{Gemmer}}, \bibnamefont{and}
  \bibinfo{author}{\bibfnamefont{G.}~\bibnamefont{Mahler}},
  \bibinfo{journal}{Eur. Phys. J. B} \textbf{\bibinfo{volume}{42}},
  \bibinfo{pages}{555} (\bibinfo{year}{2004}).

\bibitem[{\citenamefont{Michel}(2006)}]{Michel2006III}
\bibinfo{author}{\bibfnamefont{M.}~\bibnamefont{Michel}}, Ph.D. thesis,
  \bibinfo{school}{Universit\"at Stuttgart}, \bibinfo{address}{Stuttgart}
  (\bibinfo{year}{2006}).

\bibitem[{\citenamefont{Einstein}(1905)}]{Einstein1905}
\bibinfo{author}{\bibfnamefont{A.}~\bibnamefont{Einstein}},
  \bibinfo{journal}{Ann. Phys. (Leipzig)} \textbf{\bibinfo{volume}{17}},
  \bibinfo{pages}{549} (\bibinfo{year}{1905}).

\bibitem[{\citenamefont{Einstein}(1908)}]{Einstein1908}
\bibinfo{author}{\bibfnamefont{A.}~\bibnamefont{Einstein}},
  \bibinfo{journal}{Z. Elektrochem.} \textbf{\bibinfo{volume}{14}},
  \bibinfo{pages}{235} (\bibinfo{year}{1908}).

\bibitem[{\citenamefont{Gemmer and Mahler}(2003)}]{Gemmer2003}
\bibinfo{author}{\bibfnamefont{J.}~\bibnamefont{Gemmer}} \bibnamefont{and}
  \bibinfo{author}{\bibfnamefont{G.}~\bibnamefont{Mahler}},
  \bibinfo{journal}{Eur. Phys. J. B} \textbf{\bibinfo{volume}{31}},
  \bibinfo{pages}{249} (\bibinfo{year}{2003}).

\bibitem[{\citenamefont{Gemmer and Michel}(2005)}]{Gemmer2005I}
\bibinfo{author}{\bibfnamefont{J.}~\bibnamefont{Gemmer}} \bibnamefont{and}
  \bibinfo{author}{\bibfnamefont{M.}~\bibnamefont{Michel}},
  \bibinfo{journal}{Physica E} \textbf{\bibinfo{volume}{29}},
  \bibinfo{pages}{136} (\bibinfo{year}{2005}).

\bibitem[{\citenamefont{Michel et~al.}(2006)\citenamefont{Michel, Gemmer, and
  Mahler}}]{Michel2006I}
\bibinfo{author}{\bibfnamefont{M.}~\bibnamefont{Michel}},
  \bibinfo{author}{\bibfnamefont{J.}~\bibnamefont{Gemmer}}, \bibnamefont{and}
  \bibinfo{author}{\bibfnamefont{G.}~\bibnamefont{Mahler}},
  \bibinfo{journal}{Phys. Rev. E} \textbf{\bibinfo{volume}{73}},
  \bibinfo{pages}{016101} (\bibinfo{year}{2006}).

\bibitem[{\citenamefont{Michel et~al.}(2005{\natexlab{b}})\citenamefont{Michel,
  Mahler, and Gemmer}}]{Michel2005II}
\bibinfo{author}{\bibfnamefont{M.}~\bibnamefont{Michel}},
  \bibinfo{author}{\bibfnamefont{G.}~\bibnamefont{Mahler}}, \bibnamefont{and}
  \bibinfo{author}{\bibfnamefont{J.}~\bibnamefont{Gemmer}},
  \bibinfo{journal}{Phys. Rev. Lett.} \textbf{\bibinfo{volume}{95}},
  \bibinfo{pages}{180602} (\bibinfo{year}{2005}{\natexlab{b}}).

\bibitem[{\citenamefont{Gemmer and Michel}(2006)}]{Gemmer2006I}
\bibinfo{author}{\bibfnamefont{J.}~\bibnamefont{Gemmer}} \bibnamefont{and}
  \bibinfo{author}{\bibfnamefont{M.}~\bibnamefont{Michel}},
  \bibinfo{journal}{Europhys. Lett.} \textbf{\bibinfo{volume}{73}},
  \bibinfo{pages}{1} (\bibinfo{year}{2006}).

\bibitem[{\citenamefont{Borowski et~al.}(2003)\citenamefont{Borowski, Gemmer,
  and Mahler}}]{BorowskiGemmer2003}
\bibinfo{author}{\bibfnamefont{P.}~\bibnamefont{Borowski}},
  \bibinfo{author}{\bibfnamefont{J.}~\bibnamefont{Gemmer}}, \bibnamefont{and}
  \bibinfo{author}{\bibfnamefont{G.}~\bibnamefont{Mahler}},
  \bibinfo{journal}{Eur. Phys. J. B} \textbf{\bibinfo{volume}{35}},
  \bibinfo{pages}{255} (\bibinfo{year}{2003}).

\end{thebibliography}
\end{document}